\documentclass[a4paper,11pt]{article}
\usepackage{jinstpub} %

\usepackage{siunitx}
\usepackage{booktabs}
\usepackage{xspace}
\usepackage{subcaption}
\usepackage{graphicx}
\usepackage{color}
\usepackage{bm}
\usepackage{soul}

\newcommand{\geant}{\textsc{Geant4}\xspace}
\newcommand\norm[1]{\left\lVert#1\right\rVert}

\definecolor{airforceblue}{rgb}{0.36, 0.54, 0.66}
\definecolor{darkpurple}{rgb}{0.5, 0.2, 0.8}
\definecolor{dartmouthgreen}{rgb}{0.05, 0.5, 0.06}
\definecolor{taborange}{rgb}{1.0, 0.49, 0.054}
\definecolor{amaranth}{rgb}{0.9, 0.17, 0.31}

\title{CaloClouds: Fast Geometry-Independent Highly-Granular Calorimeter Simulation}

\author[a]{Erik Buhmann,}
\author[a,b]{Sascha Diefenbacher,}
\author[c]{Engin Eren,}
\author[c,d]{Frank Gaede,}
\author[a,d]{Gregor Kasicezka,}
\author[c,1]{Anatolii Korol,\note{Corresponding author.}}
\author[a]{William Korcari,}
\author[c]{Katja Kr\"uger}
\author[c]{and Peter McKeown}

\affiliation[a]{Institut f\"ur Experimentalphysik, Universit\"at Hamburg,\\Luruper Chaussee~149, 22607 Hamburg, Germany}
\affiliation[b]{Physics Division, Lawrence Berkeley National Laboratory,\\1 Cyclotron Rd, Berkeley, CA 94720, USA} 
\affiliation[c]{Deutsches Elektronen-Synchrotron DESY,\\Notkestr. 85, 22607 Hamburg, Germany} 
\affiliation[d]{Center for Data and Computing in Natural Sciences CDCS, Deutsches Elektronen-Synchrotron DESY,\\Notkestr. 85, 22607 Hamburg, Germany}

\emailAdd{anatolii.korol@desy.de}

\abstract{
Simulating showers of particles in highly-granular detectors is a key frontier in the application of machine learning to particle physics. 
Achieving high accuracy and speed with generative machine learning models would enable them to augment traditional simulations and alleviate a major computing constraint.
This work achieves a major breakthrough in this task by, for the first time, directly generating a point cloud of a few thousand space points with energy depositions in the detector in 3D space without relying on a fixed-grid structure.  
This is made possible by two key innovations: 
i) Using recent improvements in generative modeling we apply a diffusion model to generate photon showers as high-cardinality point clouds. 
ii) These point clouds of up to $6,000$ space points are largely geometry-independent as they are down-sampled from initial even higher-resolution point clouds of up to $40,000$ so-called \geant steps. 
We showcase the performance of this approach using the specific example of simulating photon showers in the planned electromagnetic calorimeter of the International Large Detector (ILD) and achieve overall good modeling of physically relevant distributions.
}


\arxivnumber{2305.04847} %

\begin{document}
\maketitle
\flushbottom

\section{Introduction}
\label{sec:Intro}

Large-scale experiments in high energy physics (HEP) have one unifying feature: the need to record and analyze an ever-increasing amount of data
produced primarily through higher collider luminosities %
and higher granularity detectors with growing numbers of readout channels.
The reduced experimental uncertainty that comes with this increased amount of higher-resolution measurement data presents one of the most promising avenues to increase the precision of measurements and the sensitivity of searches for new physical phenomena beyond the Standard Model.

In order to compare experimental measurement and theoretical prediction, modern HEP experiments rely on high-precision Monte Carlo (MC) simulations. These simulations model collider events from the initial hard scattering, over hadronization, to the detailed responses of the tracking-, calorimeter-, and muon-systems of the detector. Such detailed modeling is a highly resource-intensive process. Further, for subsequent comparison to collision data to be valid, the number of simulated events has to at least equal the number of recorded collisions. The combination of these two factors means that MC simulation alone places a significant strain on the available HEP computing resources, which, for future high-luminosity experiments, is only bound to increase further~\cite{HEPSoftwareFoundation:2017ggl,LHCC_HL:2022}.

The most notable bottleneck within the MC simulation chain is the step that has to deal with a large number of particles interacting with complex detectors --- shower simulation --- as the modeling of all involved particles is inherently time-consuming~\cite{ATLAS:2010arf}. This provides a strong incentive to find new ways of speeding up MC simulations. One of the most promising current fast simulation approaches is provided by generative machine learning (ML) models. These models can learn an underlying distribution from a given data set and can subsequently be used to generate new data from the learned distribution, allowing them to generate simulated data potentially orders of magnitude faster than classical simulations. 

The concept of generative calorimeter simulation is well-proven at this point. Its underlying mathematical viability has been demonstrated~\cite{ganplify, Calomplification} and direct applications have been shown using generative adversarial networks~\cite{CaloGan1, CaloGan2, CaloGan3, ErdmannWGAN1, ErdmannWGAN2, Vallecorsa1, Vallecorsa2,  DetectorSim2, AtlasCaloFastsim, Hashemi:2023ruu}, autoencoder-variants~\cite{gettinghigh, hadrons, Diefenbacher:2023angles}, normalizing flows~\cite{krause2021caloflow, krause2021caloflow2, Diefenbacher:2023prl} and diffusion models~\cite{Mikuni:2022xry,Mikuni:2023dvk,leigh2023pcjedi}. 
They have been successfully deployed e.g. by the ATLAS collaboration~\cite{AtlFast3}.

Previous generative calorimeter models were influenced
by machine learning models used in computer vision for image generation, therefore focused on fixed geometrical structures, either in the form of 3-dimensional tensors that mirror the design of a calorimeter or in the form of an ordered list, that connects one output to one calorimeter cell. Within the fixed-structure framework, applications have started out  using comparatively simplistic calorimeter geometries and gradually moved to higher and higher detector granularity. This directly follows the advances in calorimeter development, which similarly aim for increasingly granular detectors~\cite{CMS_hgcal, ILC-TDR}. 

It is, however, this move to higher granularity, that presents the largest challenge for fixed-structure generative models, as higher data dimensionality entails more resource-intensive generation. Most notably, the increased sparsity of highly granular detector data means a large fraction of the model evaluation is \textit{wasted} on empty cells. In an ideal scenario, this wasted computation could be avoided and only the non-empty sections of the calorimeter would be simulated.

To overcome this issue, we introduce \textsc{CaloClouds}, a generative shower simulation approach that does not rely on a fixed structure, but instead generates geometry-independent point clouds.
\footnote{The code is available at \url{https://github.com/FLC-QU-hep/CaloClouds}.}
This approach removes the need to simulate empty detector regions. Further, the produced point clouds can be directly projected into an arbitrary detector geometry, thereby allowing for effortless translation invariance and making the handling of e.g. non-regular grids or hexagonal structures straightforward.

Point cloud-based generative models have previously been explored in HEP, most notably for fast hadronization simulation~\cite{Buhmann:2023pmh, kansal2022particle, leigh2023pcjedi}, however, this work is the first to successfully expand this technique to handle the significantly larger point clouds of several thousand space points required for realistic calorimeter simulations. 

In the following, we describe the data in Section~\ref{sec:Data}, introduce the model in Section~\ref{sec:Model}, show results in Section~\ref{sec:Results}, and offer conclusions in Section~\ref{sec:Conclusion}.

\section{Data Samples}
\label{sec:Data}

We investigate the performance of our generative 
point cloud model with a purpose-built dataset that describes the energy depositions of photon showers in the electromagnetic calorimeter (ECAL) of the International Large Detector (ILD)~\cite{ILD-IDR}. 
The calorimeter data training set consists of 524k 
\footnote{The training data is available at \url{https://zenodo.org/records/10044175}.}
photon showers, with incident energy uniformly distributed between 10~GeV and 90~GeV, simulated in the ECAL of the ILD. The ILD ECAL is a highly granular sampling calorimeter, consisting of 30 layers with alternating passive tungsten absorbers and active silicon sensors. The first (as encountered by the shower) 20 absorber layers have a thickness of 2.1~mm, while the last 10 absorbers have a thickness of 4.2~mm. Every silicon layer has a thickness of 0.5~mm and subdivided into $5\,\textrm{mm}\times5\,\textrm{mm}$ cells, which are read out individually. The showers are simulated using \geant Version 10.4 (with QGSP BERT
physics list) in a detailed simulation model of the ILD detector implemented in the iLCSoft~\cite{ilcsoft} framework. In the case of the ILD ECAL, the model implemented in \textsc{DD4hep}~\cite{dd4hep} includes realistic gaps between silicon sensors and staggering between layers, such that the cells form an irregular (position-dependent) grid in 3D space. In order to describe the showers, we introduce two coordinate systems: a compact area near the photon's impact position within the calorimeter is denoted by the coordinates $[X, Y, Z]$, with $X$ and $Y$ pointing parallel to the orientation of the calorimeter layers and $Z$ pointing perpendicular to the calorimeter layers into the detector along the direction of the shower propagation and the $[X^{\prime}, Y^{\prime}, Z^{\prime}]$ coordinates that constitute the global coordinate system used in ILD, with $Z^{\prime}$ pointing parallel to the beam pipe, $X^\prime$ laying in the horizontal plane and $Y^{\prime}$ pointing vertically upwards.

The simulated photons are produced at $[X^{\prime}=0, Y^{\prime}=1811.3~\textrm{mm}, Z^{\prime}=40~\textrm{mm}]$ with a trajectory along $Y^{\prime}$, i.e. they are created right at the front of the calorimeter in order to avoid interactions before entering the calorimeter, as well as being positioned so as to avoid cracks in the detector.

During the full simulation with \geant, a very large number of individual energy depositions (on average 20,000 per shower) from secondary particles traversing the sensors is created in the sensitive materials and then added up in the actual calorimeter cells of $5\,\textrm{mm}\times5\,\textrm{mm}$ transversal size. While in principle these
steps\footnote{Following the standard \geant~\cite{g4} terminology.} could be used as input to the point cloud model, their number is prohibitively large, and therefore a pre-clustering procedure is applied before the actual training.
However, this clustering still needs to be more granular than the physical cell size to allow a shower projection into any part of the calorimeter (except changing its depth) without introducing reconstruction artefacts due to cell staggering and gaps. 

For this pre-clustering procedure, the steps are grouped by their layer number and then projected into a
regular grid with a 36 times higher granularity than that of the actual calorimeter, i.e. into grid cells of 
$0.83\,\textrm{mm}\times0.83\,\textrm{mm}$ transversal size, thereby reducing the total number of space points by a factor of circa 7.
The effects of this clustering are largely negligible when considering complete showers, as is shown in detail in \ref{App:Clustering}. 
Note, that only active steps are simulated, maintaining the main advantage of point cloud-based simulation.

\begin{figure*}[h]
    \centering
    \includegraphics[width=1\textwidth]{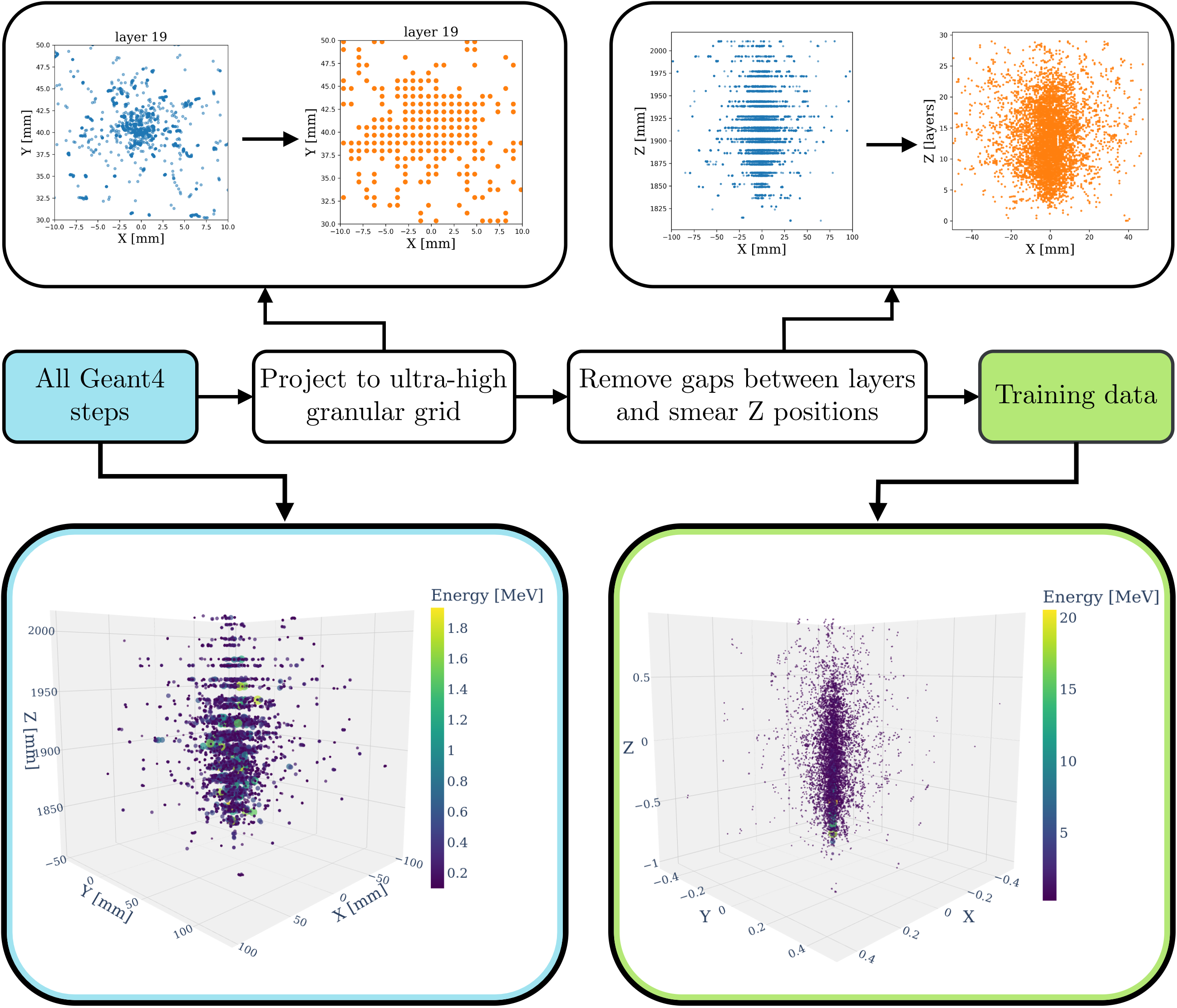}
    \caption{Illustration of the data preprocessing pipeline.
    }
    \label{fig:Data_Prepocessing}
    \hspace{0.5cm}
\end{figure*}

In order to normalize the range of inputs to the model, we define a bounding box around the showers. This box extends from $-200\,\textrm{mm}$ to $200\,\textrm{mm}$ in $X$ and $Y$ around the fixed impact point and spans across all layers, making it around seven times more voluminous than what was used in previous ILD ECAL photon shower data sets~\cite{gettinghigh}. Any clusters that still end up outside the bounding box 
--- less than 4 per mille ---  are discarded.
Finally, as there can be no recorded hits in the absorber layers, the data is transformed such that the absorber regions are removed and the active layers become contiguous. Furthermore, the $Z$ positions within the active layers are uniformly smeared within a layer to produce data that is smooth in $Z$, whereas in the ILD simulation, the energy is deposited for purely technical reasons predominantly at the center of a layer.
The cluster positions are then normalized such that the opposing boundaries of the bounding box correspond to a position of $-1$ and $1$ respectively. Figure~\ref{fig:Data_Prepocessing} presents an overview of the data preprocessing pipeline.

In addition to the training data, two validation data sets and one test set were simulated. The two validation data sets were simulated at the same position as the training data, the first containing 40k samples with energies uniformly distributed between 10~GeV and 90~GeV, and the second consisting of fixed energy photon showers with incident particle energies of 10~GeV, 50~GeV, and 90~GeV. Each set with fixed energy contains 2k showers, totaling 6k data samples. Finally, the test set contains 2k samples with fixed incident particle energy of 50~GeV that differs in the position at which the photons were simulated. The validation sets have the photons produced at  $[X^{\prime}=0, Y^{\prime}=1811.3~\textrm{mm}, Z^{\prime}=40~\textrm{mm}]$, matching the training data, while the test set places the origin of the photons at $[X^{\prime}=37.5~\textrm{mm}, Y^{\prime}=1811.3~\textrm{mm}, Z^{\prime}=-36.1~\textrm{mm}]$. This allows us to investigate the behavior of the generated point clouds under a translation to another position in the calorimeter, where the local cell geometry is different (see Section~\ref{sec:Results_Translation}).

Finally, Table~\ref{tab:reps} provides an overview of the types of point clouds considered in this work.

\begin{table}[]
\begin{tabular}{lcl}
\toprule
                          & points / shower & Note                                        \\ \midrule
All \geant steps          & 40 000                                & Initial output of \geant \\
Clustered \geant steps    & 6 000                                 & Input/output of \textsc{CaloClouds}              \\
Hits in calorimeter grid & 1 000                                 & Calculation of physics observables \\
\bottomrule
\end{tabular}
\caption{Overview of the three different types of point clouds considered in this work, indicating their role. Note that the number of points per shower (second column) only provides an order of magnitude. \label{tab:reps}}
\end{table}

\section{Model}
\label{sec:Model}

\begin{figure}[tbp]
\centering
    \begin{subfigure}[b]{0.7\textwidth}
         \centering
         \includegraphics[scale=0.25]{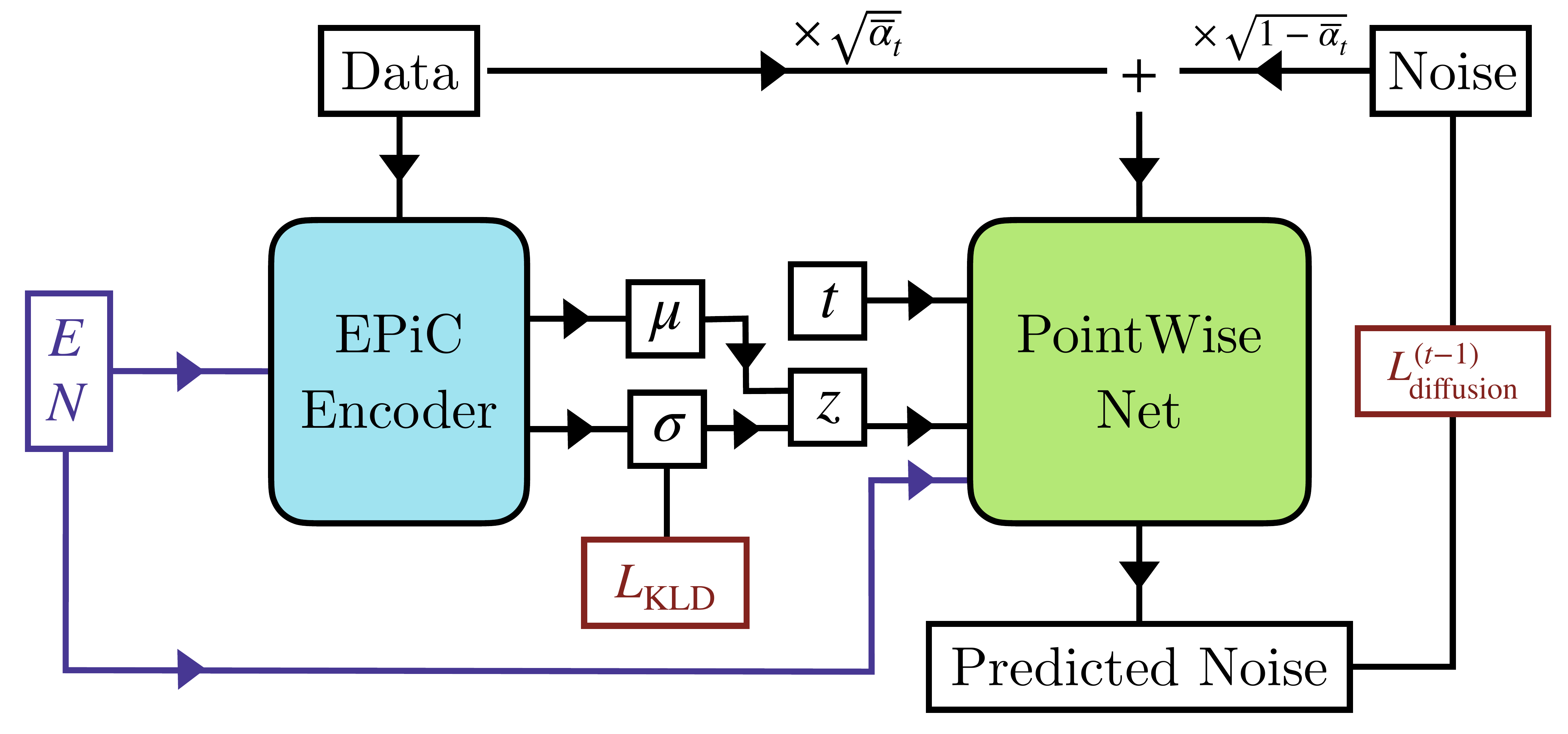}
         \caption{Training at random time step $t$}
         \label{fig:training}
    \end{subfigure}
        
    \vspace{0.5cm}

    \begin{subfigure}[b]{0.9\textwidth}
         \centering
         \includegraphics[scale=0.25]{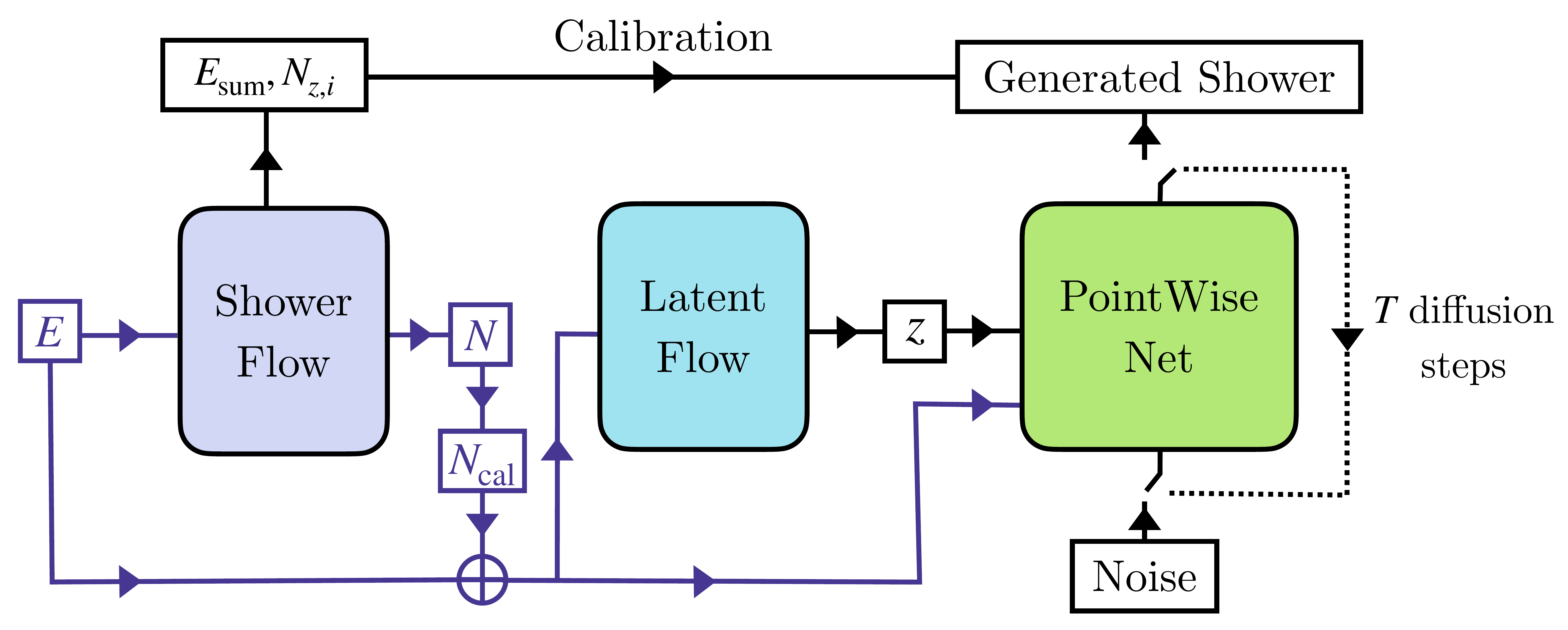}
         \caption{Sampling with reverse diffusion through all time steps $T$}
         \label{fig:sampling}
    \end{subfigure}
\caption{
Illustration of the training and sampling procedure of the \textsc{CaloClouds} architecture.  The separate training of the Shower Flow and the Latent Flow is not shown.
}
\label{fig:training_sampling_diagram}
\end{figure}

Point cloud generative models take a given number of random noise points and transform them into a point cloud of the desired shape (potentially with additional point features as in our case). This makes producing clouds with a variable number of points inherently challenging. To address this cardinality issue, we employ a two-step approach during model training and a three-step approach during sampling, drawing inspiration from Ref.~\cite{luo2021diffusion}. 
In the following, we first briefly outline the overall structure with additional details provided in corresponding subsections.
An illustration of the training and sampling procedure is shown in Figure~\ref{fig:training_sampling_diagram}. 

During training, the original data is first encoded with a Variational-Autencoder (VAE)-like \textit{EPiC Encoder}, conditioned on the incident particle energy and the number of points, into a near-Gaussian distributed latent space. Second, this latent space, together with the incident energy and number of points, is used in a conditional point cloud diffusion model termed \textit{PointWise Net}.

During sampling, the encoded latent space is generated with a conditional \textit{Latent Flow} model. 
Since this Latent Flow needs to be conditioned on the incident energy and the number of points, a second \textit{Shower Flow} is employed during sampling to generate an appropriate number of points from a requested incident energy. 
This way, the only conditional variable for the whole model is the particle incident energy $E$.
Additionally, the Shower Flow generates the total visible energy of the calorimeter point cloud $E_\mathrm{sum}$ as well as the number of points per layer $N_{z,i}$ for a post-diffusion calibration of the generated point cloud.

In Section~\ref{sec:Models_Diffusion} we discuss first the PointWise Net at the core of the overall proposed \textsc{CaloClouds} architecture. 
Then, in Section~\ref{sec:encoder_latent_flow}, we introduce the EPiC Encoder used during training and the Latent Flow used to perform sampling. 
Finally, in Section~\ref{sec:shower_flow} we outline the Shower Flow used to generate the conditioning variables as well as calibration factors.

\subsection{Diffusion Point Cloud Generator}
\label{sec:Models_Diffusion}

\begin{figure}[tbp]
\centering
    \begin{subfigure}[b]{0.45\textwidth}
         \centering
         \includegraphics[scale=0.2]{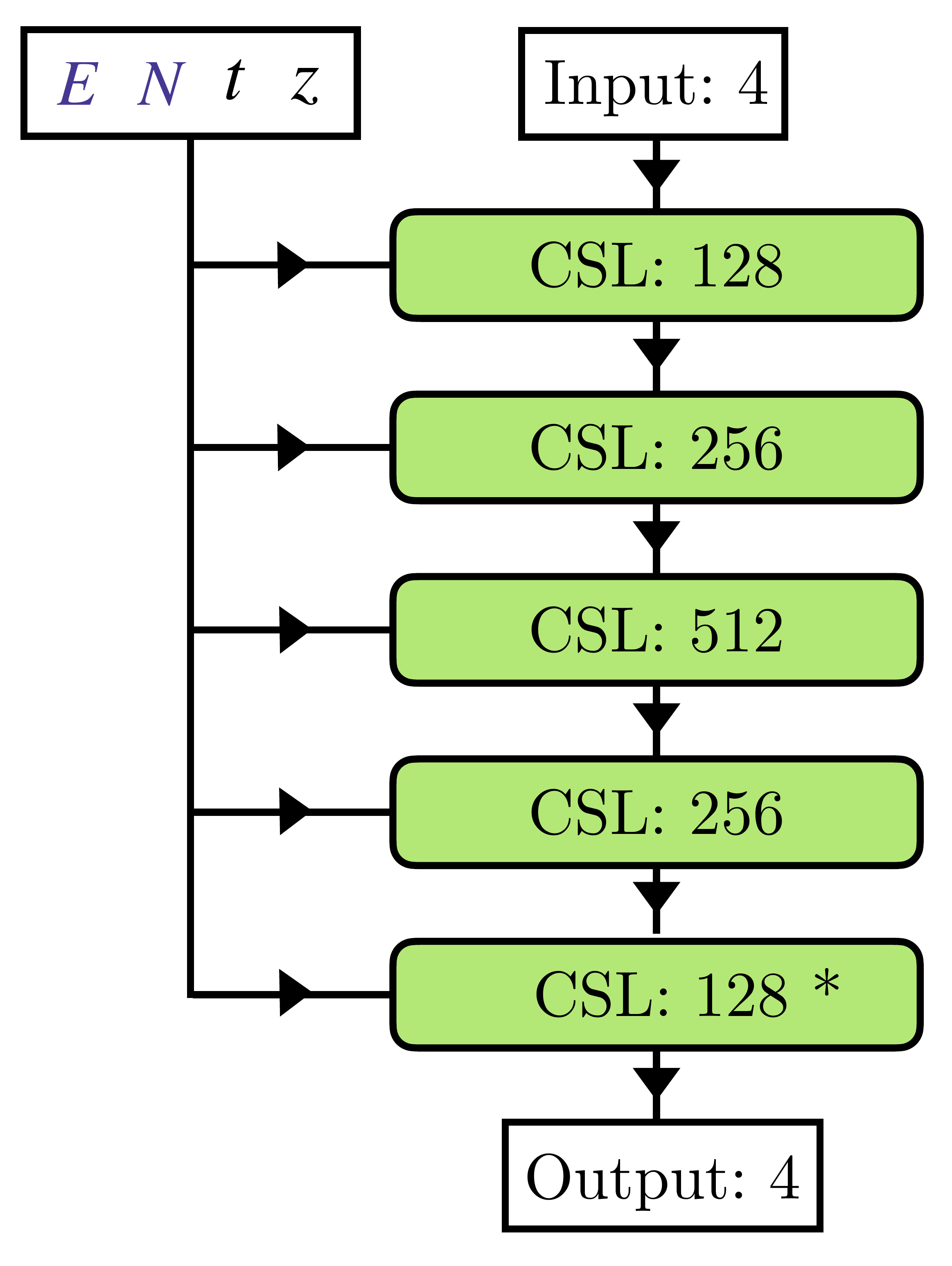}
         \caption{PointWise Net}
         \label{fig:pointwisenet}
    \end{subfigure}
    \begin{subfigure}[b]{0.45\textwidth}
         \centering
         \includegraphics[scale=0.2]{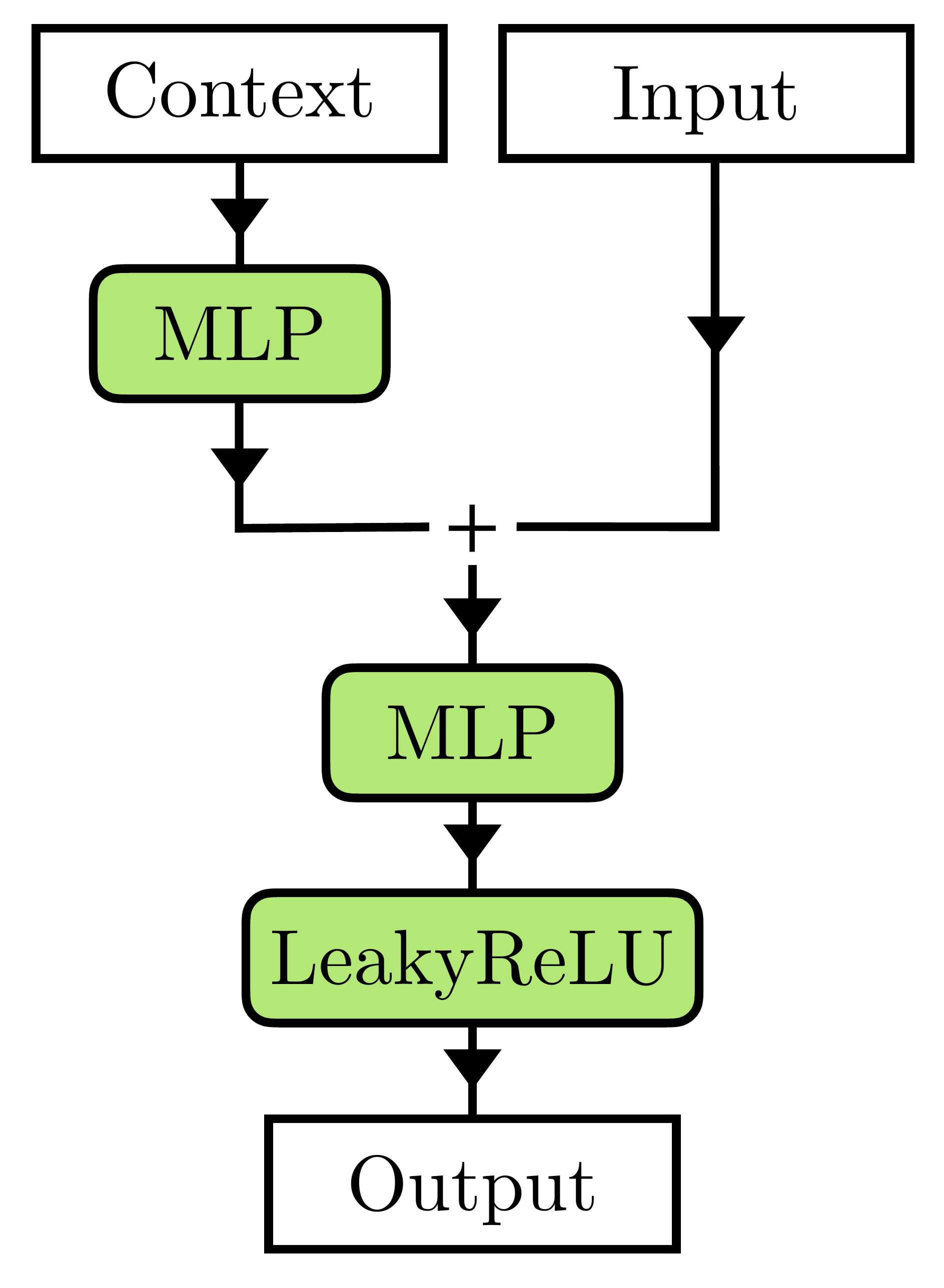}
         \caption{ConcatSquash Layer (CSL)}
         \label{fig:concatsquashlinear}
    \end{subfigure}
\caption{
Illustration of the \textsc{CaloClouds}' PointWise Net (a) consisting of multiple ConcatSquash layers (b). The number of hidden dimensions is indicated. MLP denotes a multi-layer perceptron. $^*$No activation function is applied in the last layer of the PointWise Net. 
}
\label{fig:diffusion_generator}
\end{figure}

The main part of the \textsc{CaloClouds} architecture is the PointWise Net which is trained as a diffusion probabilistic model and was introduced in Ref.~\cite{luo2021diffusion}. 
In general, diffusion models~\cite{sohldickstein2015deep} learn a Markov chain of Gaussian distortions that over $T$ time steps diffuse data to noise. 
Here the data is a point cloud $X^{(0)} = \{ \bm{x}_i^{(0)} \}_{i=1}^N$ comprised of $N$ points with the exponent in parentheses denoting the time step.

During training, a forward diffusion process for every point is modeled as a Markov chain:
\begin{align}
\begin{split}
    q( \bm{x}_i^{1:T} | \bm{x}_i^{(0)} ) &= \prod_{t=1}^T q( \bm{x}_i^{(t)} | \bm{x}_i^{(t-1)}), 
    \\
    q( \bm{x}_i^{(t)} | \bm{x}_i^{(t-1)} ) &= \mathcal{N}( \bm{x}_i^{(t)} | \sqrt{ 1 - \beta_t} \bm{x}_i^{(t-1)}, \beta_t \bm{I} ),
\end{split}
\end{align}
for time steps $t \in 1, \ldots, T$,  $\mathcal{N}$ denoting the normal distribution, and variance schedule $\beta_1 \cdots \beta_T$ controlling the diffusion process.
To generate realistic point clouds from a simple noise distribution, the reverse diffusion process starting from noise $p( \bm{x}_i^{(T)} ) = \mathcal{N}(\bm{0}, \bm{I})$ is defined as:
\begin{align}
\begin{split}
    p_{\bm{\theta}}( \bm{x}_i^{(0:T)} | \bm{z} )  &= p( \bm{x}_i^{(T)} ) \prod_{t=1}^T p_{\bm{\theta}} ( \bm{x}_i^{(t-1)} | \bm{x}_i^{(t)}, \bm{z}), 
    \\
    p_{\bm{\theta}} ( \bm{x}_i^{(t-1)} | \bm{x}_i^{(t)}, \bm{z}) &= \mathcal{N}( \bm{x}_i^{(t-1)} | \bm{\mu_\theta} ( \bm{x}_i^{(t)}, t, \bm{z} ), \beta_t \bm{I}),
\end{split}
\end{align}
where the estimated mean $\bm{\mu_\theta}$ is modeled by a neural network with parameters $\theta$ and the generation process is steered by a conditioning latent vector $\bm{z}$.
This latent variable is crucial as the generated point cloud needs to have consistent global properties, i.e. realistic energy and center of gravity. 
As all initial noise points $X^{(T)} = {\bm{x}_i^{(T)}}_{i=1}^N$ are sampled identical and independently distributed (i.i.d.), the latent $\bm{z}$ is the only source of shared information for the generated point cloud $X^{(0)}$.

For practical training, Ref.~\cite{ho2020denoising} showed that $\bm{x}_i^{(t)}$ can be sampled directly from data $\bm{x}_i^{(0)}$ without going through all previous steps:
\begin{align}
    q( \bm{x}_i^{(t)} | \bm{x}_i^{(0)} ) 
    = \mathcal{N}( \bm{x}_i^{(t)} | \sqrt{\overline{\alpha}_t} \bm{x}_i^{(0)}, (1-\overline{\alpha}_t)\bm{I}),
    \label{eq:xt_from_x0}
\end{align}
with notation $ \alpha_t := 1 - \beta_t$ and  $\overline{\alpha}_t := \prod_{s=1}^t \alpha_s$.
This allows parameterising  $\bm{x}_i^{(t)}$ as $\bm{x}_i^{(t)} ( \bm{x}_i^{(0)}, \epsilon) = \sqrt{\overline{\alpha}_t} \bm{x}_i^{(0)} + \sqrt{1-\overline{\alpha}_t} \epsilon$ with noise $\epsilon \sim \mathcal{N}(\bm{0},\bm{I})$.
The model approximating the mean $\bm{\mu_\theta}$ can be parameterized as: 
\begin{align}
    \bm{\mu_\theta}(\bm{x}_i^{(t)}, t) 
    = \frac{1}{\sqrt{\overline{\alpha}_t}} \left( \bm{x}_i^{(t)} - \frac{\beta_t}{\sqrt{1-\overline{\alpha}_t}} \bm{\epsilon_\theta} (\bm{x}_i^{(t)},t,\bm{z}) \right),
\end{align}
where $\bm{\epsilon_\theta} (\bm{x}_i^{(t)},t,\bm{z})$ is a neural network which predicts $\bm{\epsilon}$ from $\bm{x}_i^{(t)}$. 
Hence, the model $\bm{\mu_\theta}$ was rephrased as a noise predictor $\bm{\epsilon_\theta}$ which can be optimized with a simple L2 loss:

\begin{align}
    L_i^{(t-1)}
    = \mathbb{E}_{\bm{x}_i^{(0)},\bm{\epsilon}, \bm{z} } 
    \left[ \norm{ \bm{\epsilon} - \bm{\epsilon_\theta} ( \sqrt{\overline{\alpha}_t} \bm{x}_i^{(0)} + \sqrt{1-\overline{\alpha}_t} \bm{\epsilon}, t, \bm{z}  )}^2  \right].
\end{align}
Therefore the diffusion loss for the full point cloud at time step $t$ is given by $L_\mathrm{diffusion}^{(t-1)} = \frac{1}{N} \sum_{i=1}^N L_i^{(t-1)}$. 
For efficient training, a random time step $t \in {1, \ldots, T}$ is chosen at each optimization step. 
A full derivation of the PointWise Net and its training objective can be found in Ref.~\cite{luo2021diffusion}.
We optimize our model with a total of $T=100$ time steps with a quadratic variance scheduler between $\beta_T = 0.02$ and $\beta_1 = 10^{-4}$.

The neural network $\bm{\mu_\theta}$ is the same for each point $\bm{x}_i^{(t)} \in X^{(t)}$, therefore we can sample permutation-equivariant point clouds with a variable number of points $N$.
Output and input dimensions of the network need to be identical --- in our case four (three spatial dimensions and energy).
We utilize the same PointWise Net architecture as used in Ref.~\cite{luo2021diffusion}. 
An overview of the architecture and the layer structure is shown in Figure~\ref{fig:diffusion_generator}. 
It consists of multiple \textit{ConcatSquash} layers~\cite{grathwohl2018ffjord} which are conditioned on a global context vector --- in this case a concatenated vector of the incident particle energy $E$, the number of points $N$, the time features $t$, and the latent vector $\bm{z}$. 
The ConcatSquash layers are essentially two-layer fully connected networks, with the first hidden layer aligning the dimensionality of the context conditioning vector with the hidden dimensionality of the layer input.
A \textsc{LeakyReLU} activation follows the second hidden layer. 

This architecture allows for each point to be sampled  
independently of all other points of the point cloud. 
Note that the assumption of i.i.d.~sampling of each point of the calorimeter shower is theoretically not optimal as in reality, the photon shower cascade develops through particle interactions.
However, due to the simple topological structure of electromagnetic showers, we assume the points to be i.i.d. and gain encouraging generative fidelity with this approach. 
Since we do not model any particle interactions, the point cloud generation scales $\mathcal{O}(\mathrm{N})$ with the number of points, which is crucial since we generate several thousand points per shower.
We performed a few early experiments with the equivariant point cloud (EPiC) layers introduced in Ref.~\cite{Buhmann:2023pmh} as they allow for point interactions with linear computational scaling but did not observe improved performance. 
This is likely due to the EPiC layers being optimised for much smaller point clouds of $\mathcal{O}(10)$ cardinality.
Graph network and transformer approaches for point cloud generation such as Ref.\cite{lee2019set, kosiorek2020conditional, kim2021setvae, kansal2022particle, dibello2022conditional, Kansal_2023} generally scale $\mathcal{O}(\mathrm{N}^2)$ and are therefore significantly slower than our model. 
However, fast implementations of attention, graph networks or sequence convolutions~\cite{GarNet_2019,dao2022flashattention,gu2022efficiently,poli2023hyena} might be interesting avenues to explore in the future, e.g. for generating hadronic showers including nuclear interactions.

\begin{figure*}[h]
    \centering
    \includegraphics[width=1\textwidth]{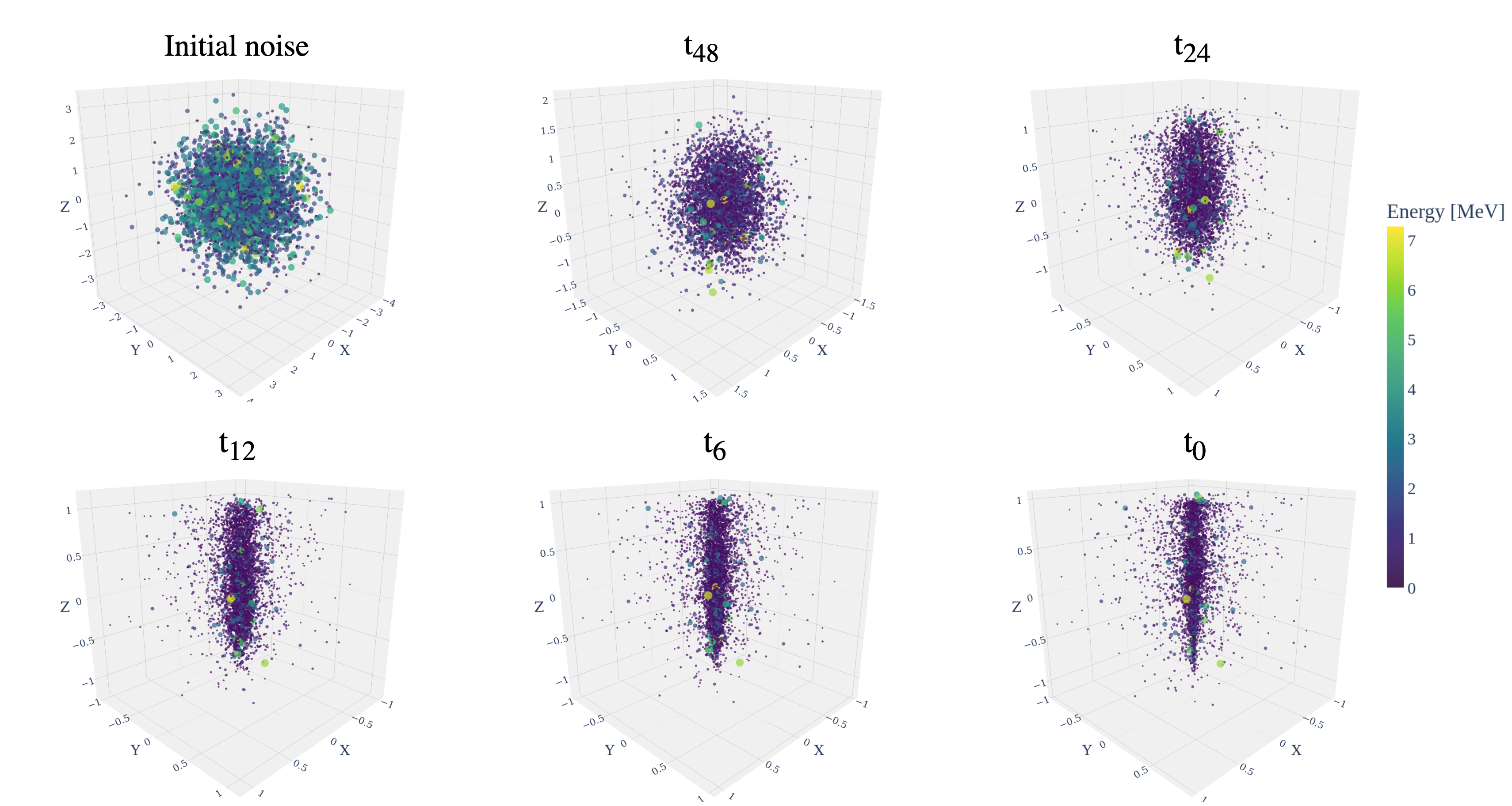}
    \caption{Illustration of the reverse diffusion process. Starting from the initial noise.  The color scale corresponds to the point energy. 
    }
    \label{fig:Sim_EnergySum}
    \hspace{0.5cm}
\end{figure*}

Figure~\ref{fig:Sim_EnergySum} shows a visualization of the calorimeter point cloud generation at different stages of the reverse diffusion process (i.e. starting with noise).

\subsection{Encoder \& Latent Flow}
\label{sec:encoder_latent_flow}

The PointWise Net generates point clouds via the reverse diffusion process. 
For the resulting point clouds to be realistic, PointWise Net needs to be conditioned on additional quantities beyond multiplicity $N$ and shower energy $E$.

In principle, one could add additional physically relevant quantities such as the total visible energy, the center of gravity, or the shower start as explicit conditioning features. 
However, such a choice of observables might bias the generated showers.
Instead, we opt for learning an additional global context vector $\bm{z}$ to capture any other relevant distributions via an additional encoder.

This encoding is learned by an Equivariant Point Cloud (EPiC) Encoder using three EPiC layers introduced in Ref.~\cite{Buhmann:2023pmh} with a hidden dimensionality of 128. 
The EPiC Encoder is conditioned on $E$ and $N$ and learns to encode the original \geant point cloud into two latent space vectors $\bm{\mu}$ and $\bm{\sigma}$. 
Similar to the encoder in a VAE, $\bm{\mu}$ and $\bm{\sigma}$ are regularised towards a Gaussian distribution with the Kullback-Leibler divergence (KLD) loss and the latent space $\bm{z}$ is sampled with the reparametrization trick~\cite{kingma_welling_variational_autoencoder}.
The KLD loss is given by:
\begin{align}
    L_\mathrm{KLD} = D_\mathrm{KL}(\mathcal{Z} || \mathcal{N}(0,1))
    = - \frac{1}{2}\left( 1 + \mathrm{log}(\bm{\sigma}^2) - \bm{\mu}^2 - \bm{\sigma}^2  \right),
\end{align}
with the latent variables sampled via $\bm{z} \sim \mathcal{Z} = \mathcal{N}(\bm{\mu}, \bm{\sigma}^2)$.
We set the size of $\bm{z}$ to 256, the default in Ref.~\cite{luo2021diffusion}, without performing a hyperparameter optimisation.

During training, the EPiC Encoder and the point cloud generator are trained in parallel using a combination of $L_\mathrm{KLD}$ and the diffusion reconstruction loss. At a randomly sampled time step $t$ this results in the total loss function:
\begin{align}
    L_\mathrm{total}^{(t-1)} = w_\mathrm{KLD} \cdot \mathrm{max}(L_\mathrm{KLD}, m_{KLD}) + L_\mathrm{diffusion}^{(t-1)},
\end{align}
with the KLD weight $w_\mathrm{KLD} = 10^{-3}$. 
To prevent a posterior collapse of $\bm{z}$ we enforced a minimum KLD loss of $m_\mathrm{KLD} = 1.0$. 
The EPiC Encoder and the PointWise Net were implemented using \textsc{PyTorch}~\cite{pytorch} and trained together for a total of 800k iterations using the \textsc{Adam}~\cite{adam} optimizer with a learning rate scheduled between $10^{-3}$ and $10^{-4}$.

During sampling, we generate the encoded latent space $\bm{z}$ with a Normalizing Flow model~\cite{normalizingflows2016} termed Latent Flow. 
This Latent Flow model is conditioned on $E$ and $N$ and consists of ten coupling blocks with monotonic rational-quadratic splines~\cite{durkan2019neural} each with two layers, a hidden dimensionality of 128, and \textsc{LeakyReLU} activations. 
The Latent Flow is trained simultaneously with the EPiC Encoder and the diffusion model via the negative log-likelihood loss but with a separate \textsc{Adam} optimizer and a learning rate scheduler.
It was implemented using the \textsc{nflows} package~\cite{nflows}.

A similar encoding and latent flow strategy are implemented in Ref.~\cite{luo2021diffusion}, however, we choose a more advanced encoder and flow architecture as well as choosing to disentangle the encoder and diffusion loss from the flow training loss to achieve a more flexible optimization regime.

\subsection{Shower Flow \& Post-Diffusion Calibration}
\label{sec:shower_flow}

To estimate the number of points $N$ in a point cloud i.e. $N_\mathrm{gen}$ for a requested incident energy $E$, we employ a separately trained normalizing flow model, called Shower Flow.
This Shower Flow is trained to generate the showers' total visible energy $E_\mathrm{sum}$ and the number of points per calorimeter layer $N_{z,i}$ for the purpose of post-diffusion calibration. 
For consistency, the total number of points $N_\mathrm{gen}$ is defined by $N_\mathrm{gen} = \sum_{i=0}^{29} N_{z,i}$.

The Shower Flow consists of ten blocks with seven coupling layers each. 
Out of these, six are based on affine transformations~\cite{RealNVP} and one is based on element-wise rational splines~\cite{durkan2019neural}. 
Each coupling layer is conditioned on the incident particle energy $E$. 
This flow was implemented with the \textsc{Pyro} package~\cite{bingham2019pyro}.

While the number of points $N_\mathrm{gen}$ generated by the Shower Flow describes the corresponding \geant distribution $N_\mathrm{data}$ well, the number of calorimeter hits, i.e. the occupancy $O_\mathrm{gen}$, of the projected point cloud may not necessarily correctly match the corresponding \geant distribution $O_\mathrm{data}$. 
This occurs as the projection of the space points into cells of the detector grid has two possible failure modes: either too many points $N_\mathrm{gen}$ are projected into the same calorimeter cell and therefore the number of hits $O_\mathrm{gen}$ is lower than the expected $O_\mathrm{data}$, or too few points are projected together leading to an overestimation of $O_\mathrm{data}$. 
For our model, we observed the latter case since for a given shower the $O_\mathrm{gen}$ is up to 10\% larger than the $O_\mathrm{data}$. To resolve this mismatch, we introduced an energy-dependent correction factor 
$c_N(N) = p_{\mathrm{gen}}(p_{\mathrm{data}}(N) ) / N$. Here, $p_{\mathrm{data}}$ represents a cubic polynomial fit of $N_\mathrm{data}$ to $O_\mathrm{data}$, while $p_{\mathrm{gen}}$ corresponds to a cubic polynomial fit of $O_\mathrm{gen}$ to $N_\mathrm{gen}$.
This correction factor is used to rescale $N_\mathrm{gen}$ predicted by the Shower Flow as well as the predicted points per layer $N_{z,i}$, resulting in a calibrated number of points $N_\mathrm{cal} = c_N \cdot N_\mathrm{gen}$ and $N_{z,i,\mathrm{cal}} = c_N \cdot N_{z,i}$. The calibrated number of points is then used for sampling from the diffusion model.

Finally, we consider an additional rescaling of the point features following the diffusion models' generation:
The post-diffusion calibration is performed in four steps.
First, in case negative point energies were generated, they are set to zero.
Second, the total energy of the point cloud is rescaled to match the predicted $E_\mathrm{sum}$.
Third, all points are ordered based on their Z-coordinate and then iteratively the first $N_{z,i=0}$ points are set to Z-coordinate index $z_i=0$, afterward the next $N_{z,i=1}$ points are set to coordinate $z_i=1$ and so on until layer index $z_i=29$.
This way, the total energy and the energy distribution in the Z-direction are well calibrated, otherwise both distributions are challenging to model with the diffusion model alone precisely. 
Fourth, a calibration of the center of gravity in the X- and Y-directions is performed by shifting all X- and Y-coordinates by their mean center of gravity of the training data set.

Overall the \textsc{CaloClouds} model together with these calibrations achieves high fidelity on a number of important shower physics observables as presented in the following section.

\section{Results}
\label{sec:Results}

In the following Section~\ref{sec:Results_Physics}, we compare the distributions produced by the \textsc{CaloCloud} model to the \geant ground-truth for several per-shower variables and global observables. 
For this, the space points generated with \textsc{CaloCloud} are projected back into the real and irregular grid-like geometry of the cells in the ILD detector model --- in the same manner as in the \geant simulation.
Section~\ref{sec:Results_Translation} then demonstrates the model's ability to generate showers originating from any arbitrary incident point of the particle along the detector surface.
Finally, in Section~\ref{sec:Results_Timings} we conclude with an investigation of our model's speed-up in comparison to \geant simulation.

\subsection{Physics Performance}
\label{sec:Results_Physics}

\begin{figure*}[h]
    \centering
    \includegraphics[width=0.32\textwidth]{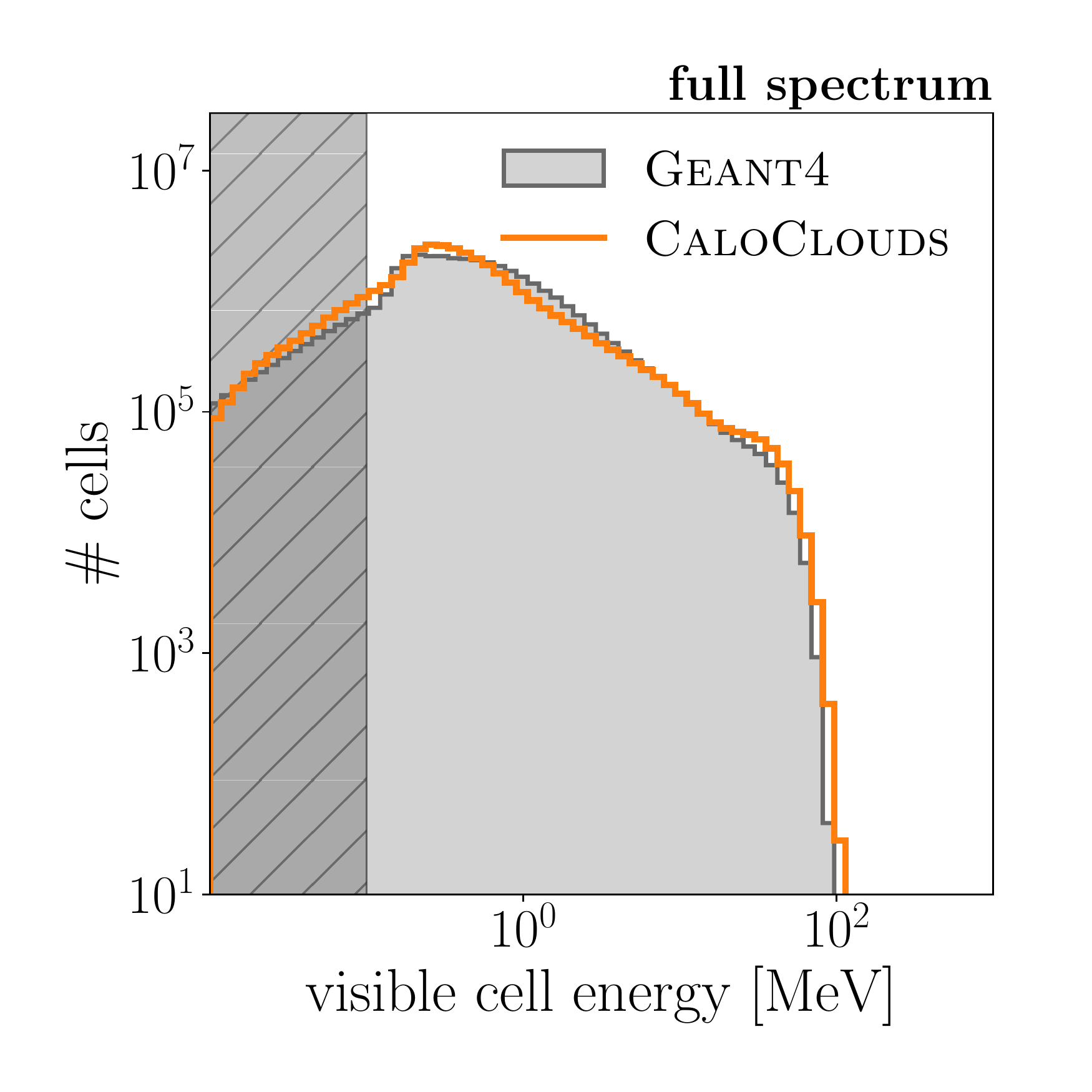}
    \includegraphics[width=0.32\textwidth]{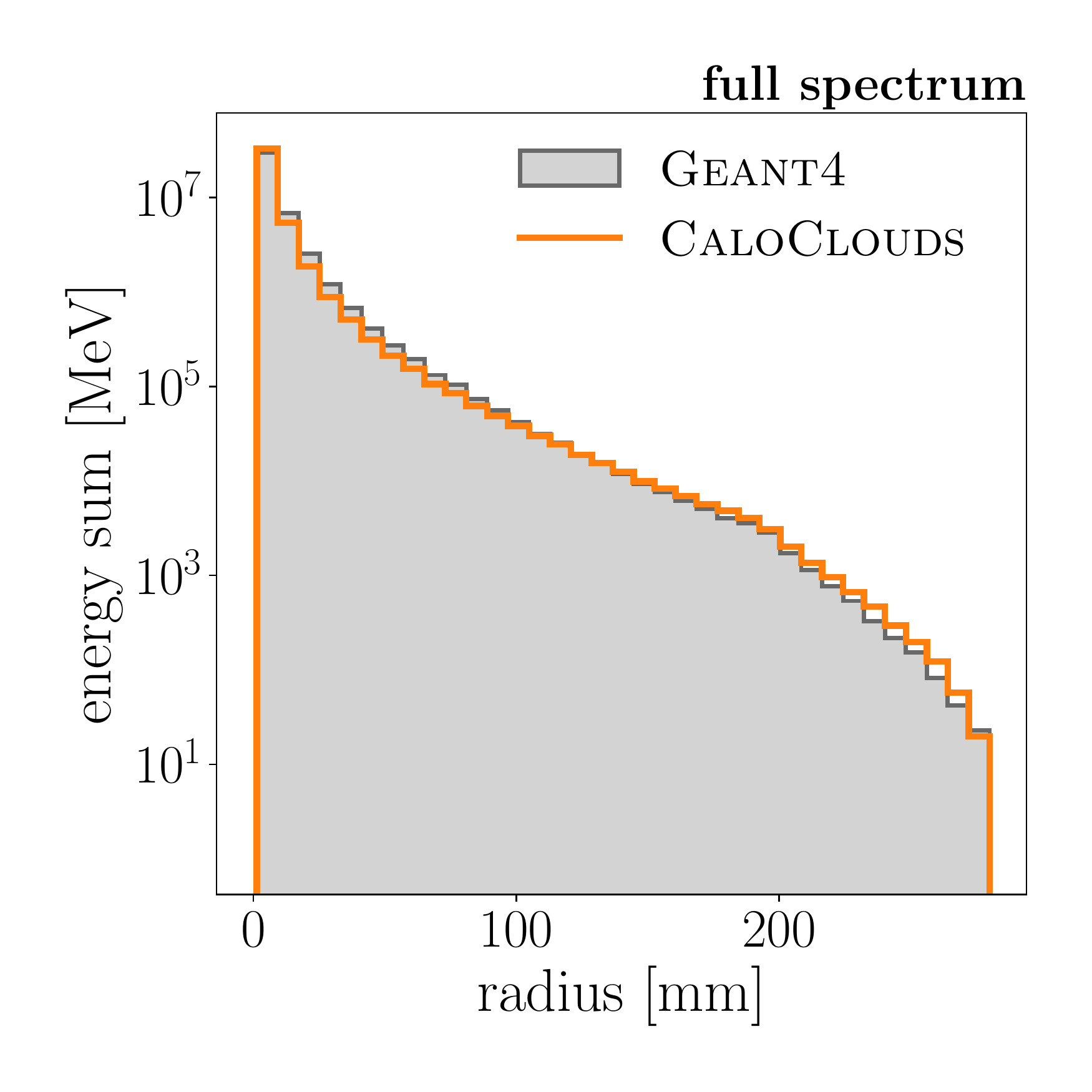}
    \includegraphics[width=0.32\textwidth]{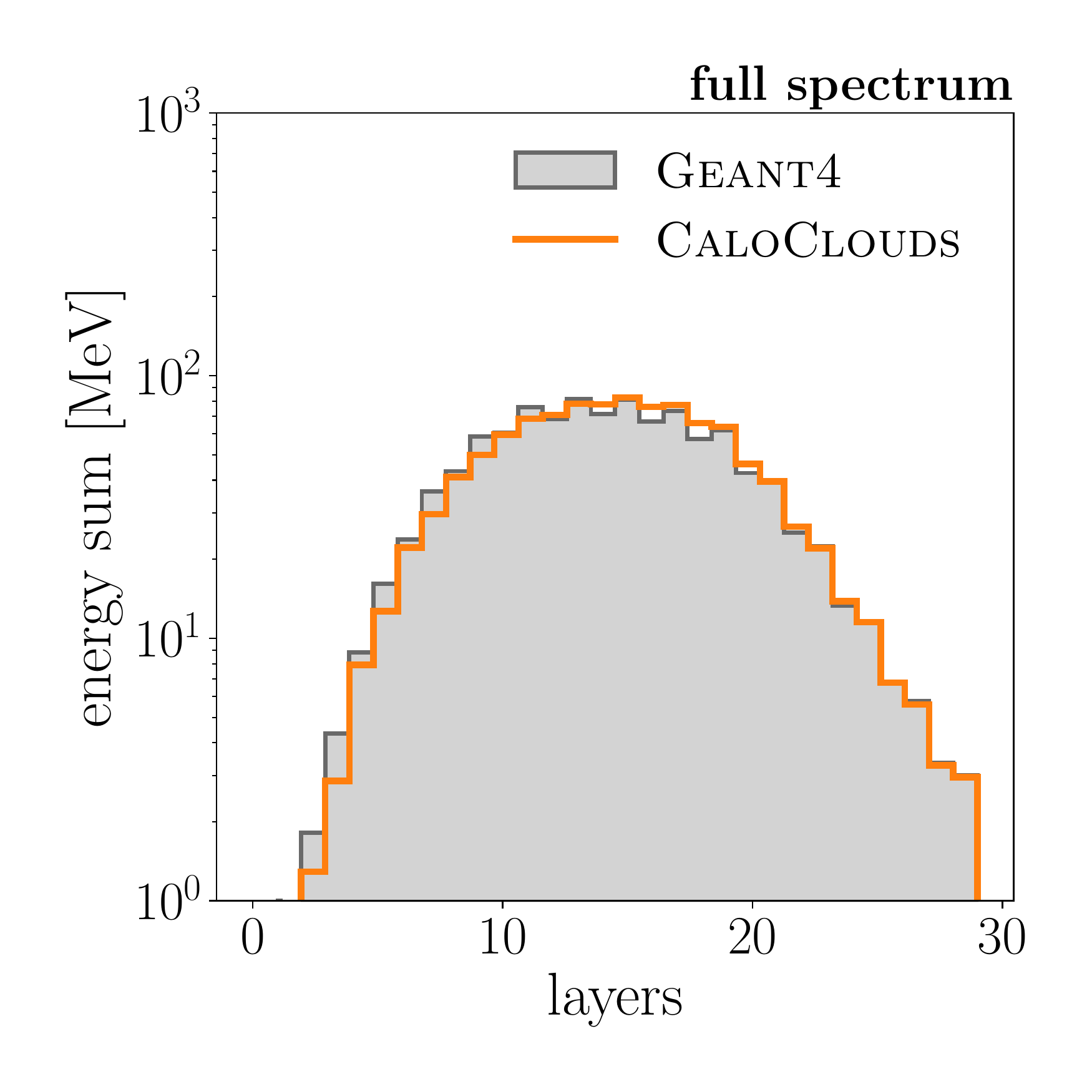}
    \caption{Histograms of the cell energies (left), radial shower profile (center), and longitudinal shower profile (right) for both \geant and \textsc{CaloClouds}. In the per-cell energy distribution, the region below 0.1~MeV is grayed out (see main text for details). All distributions are calculated for a uniform distribution of incident particle energies.
    }
    \label{fig:Ehits_Radial_Spinal}
    \hspace{0.5cm}
\end{figure*}

First, we consider how well the distribution of measured energies across a broad range of 
incident energies is described after projecting back to the physical coordinate system.
Figure~\ref{fig:Ehits_Radial_Spinal} (left) shows the per-cell energy distribution. Overall, this is well described by the diffusion model, with the largest differences occurring at regions in the spectrum where the slope changes (i.e. at 0.5~MeV and 100~MeV). 
Here, the shaded region indicates energies below 0.1 MeV, corresponding to half the energy deposited by a minimally ionizing particle (MIP). As these energies will be dominated by detector noise they are ignored in the reconstruction process and therefore are also not considered when calculating other quantities.

The radial shower profile (Figure~\ref{fig:Ehits_Radial_Spinal} (center))
and the longitudinal shower profile (Figure~\ref{fig:Ehits_Radial_Spinal} (right))
are overall well-learned by the diffusion model with some differences observed at the edges of the distribution.

\begin{figure*}[h]
    \centering
    \includegraphics[width=1\textwidth]{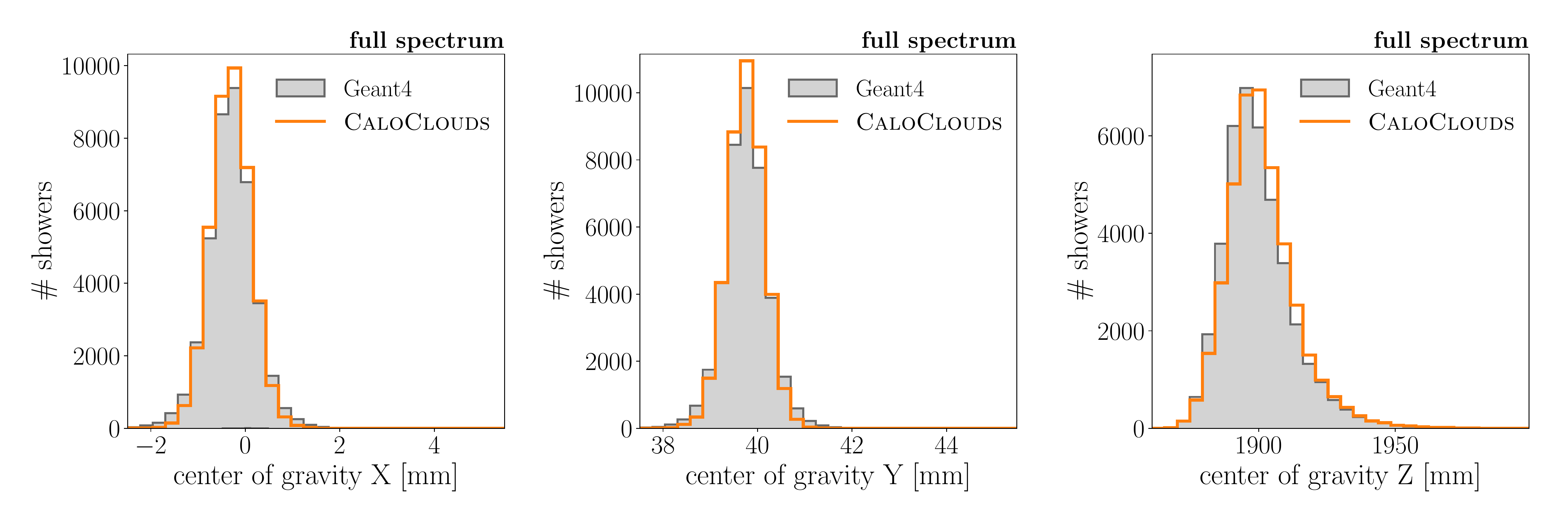}
    \caption{
    Position of the center of gravity of showers along the X (left), Y (center), and Z (right) directions. All distributions are calculated for a uniform distribution of incident particle energies.    
    }
    \label{fig:CoG}
    \hspace{0.5cm}
\end{figure*}

Next, we consider the modeling of the center of the shower in Figure~\ref{fig:CoG}. 
For this, the center of gravity (i.e. the energy-weighted centroid) is calculated per shower along the $X$ (left), $Y$ (center), and $Z$ (right) direction. The radial directions in $X$ and $Y$ are well described, with the diffusion model predicting a slightly more peaked distribution in both cases. In the longitudinal direction, the shape is captured correctly by the diffusion model as well, with the peak shifted approximately 10~mm compared to \geant.

\begin{figure*}[h]
    \centering
    \includegraphics[width=0.49\textwidth]{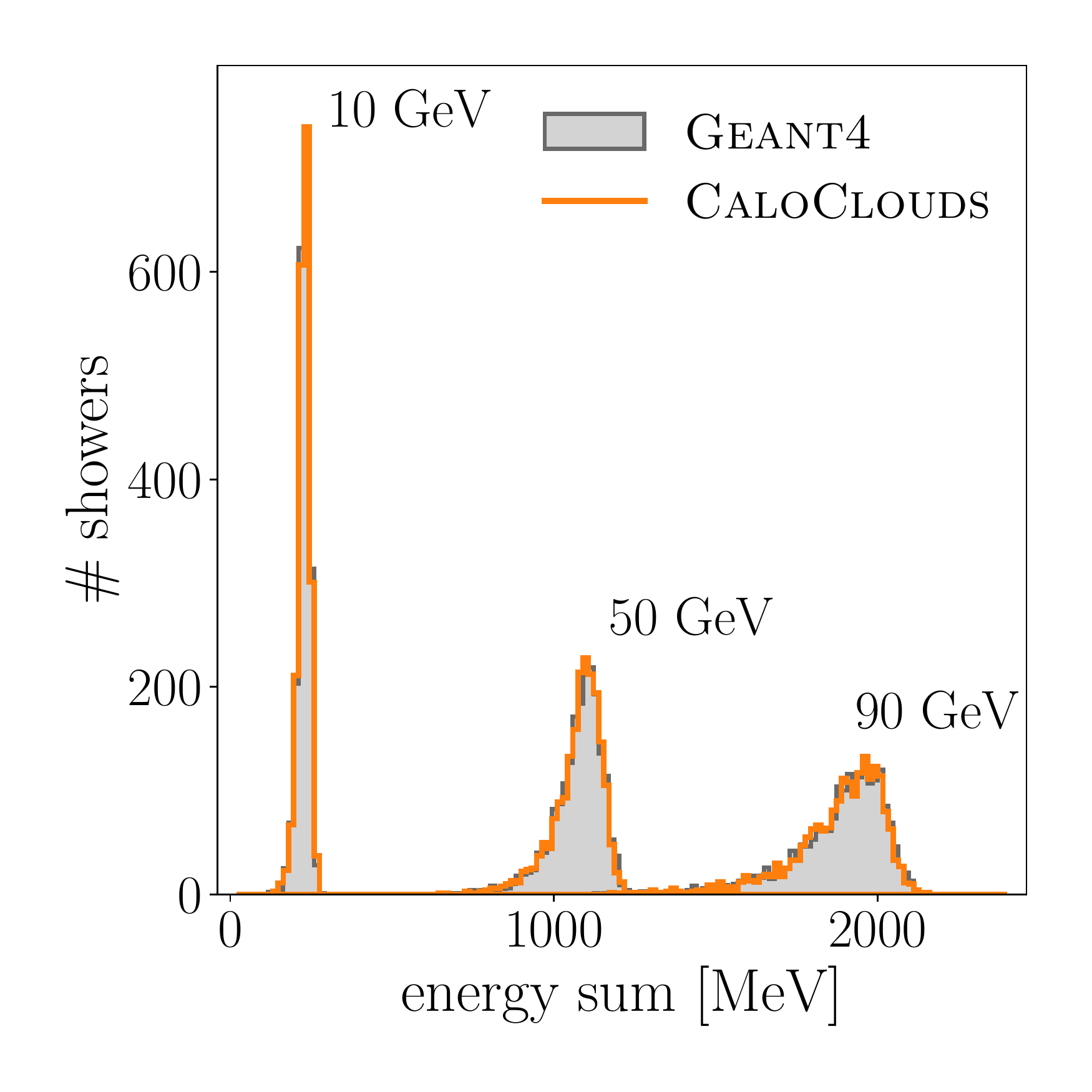}
    \includegraphics[width=0.49\textwidth]{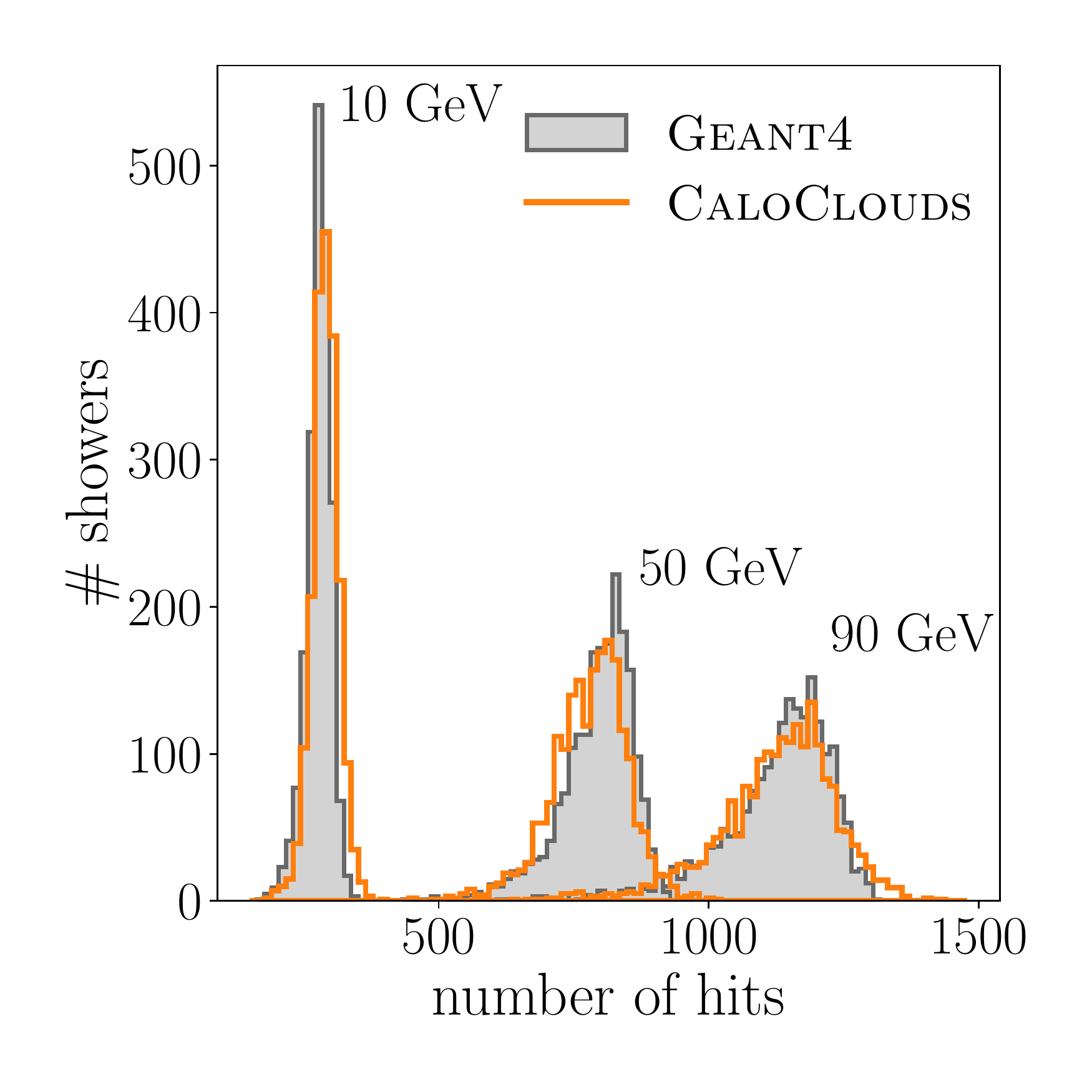}
    \caption{
    Visible energy sum (left) and the number of hits (right) distributions, for 10, 50, and 90 GeV showers.
    }
    \label{fig:Esum_Occ}
    \hspace{0.5cm}
\end{figure*}

To test whether the response for different shower energies is learned correctly, we consider the visible energy (left) and the overall number of hits above the threshold (right) in Figure~\ref{fig:Esum_Occ}. The distributions are shown for incident energies of 10, 50, and 90~GeV. At all incident energies, the visible energy distributions are very well modeled -- an important feature for a calorimeter simulation as it is the width of these distributions that directly determines the energy resolution of the detector. The distributions for the number of hits above the threshold are modeled well with, some deviations visible at higher energies.

\subsection{Shower Translation}
\label{sec:Results_Translation}

\begin{figure*}[h]
    \centering
    \includegraphics[width=0.49\textwidth]{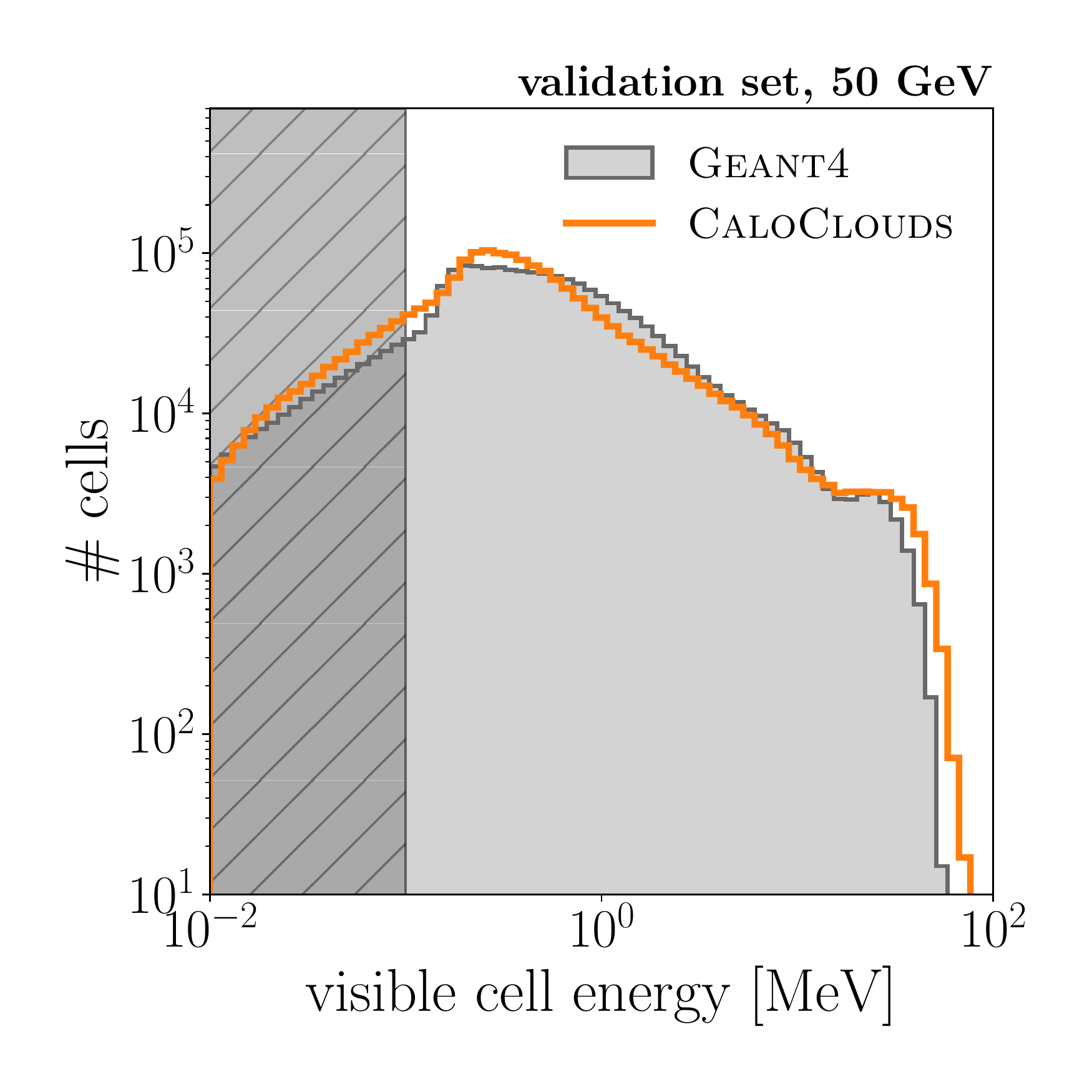}
    \includegraphics[width=0.49\textwidth]{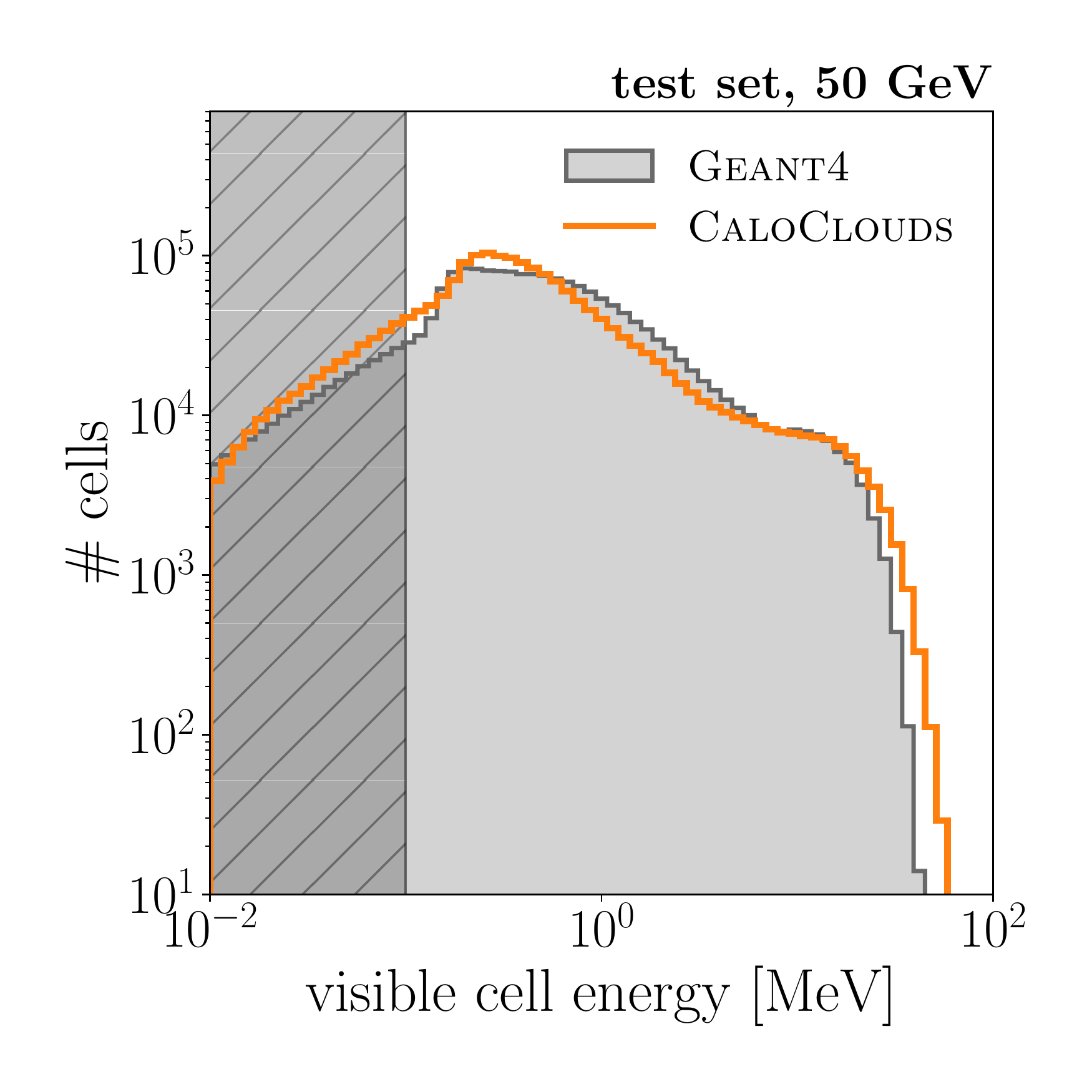}
    \caption{Per-cell energy distribution for the 50~GeV validation (left) data set, created at the same position as the training data set and for a 50~GeV test (right) data set simulated at a different position with the generated point cloud translated to this position.}
    \label{fig:hits_test}
    \hspace{0.5cm}
\end{figure*}

To test the capability of the \textsc{CaloClouds} architecture to generate photon showers originating from arbitrary incident points on the detector, we compare the model using the two different data sets introduced in Section~\ref{sec:Data}: a validation data set that has been simulated at the same position in the calorimeter as the training data set and a test data set simulated at a different position in the calorimeter. Depending on the exact impact point in the calorimeter the local cell geometry will look different, for example, due to the local staggering of cells between layers, the position of the impact point with respect to the cell centers, gaps in the cell structure at the edge of silicon sensors, etc.
The distribution that is most susceptible to potential artifacts resulting from such a translation is the cell energy distribution. This distribution is shown in Figure~\ref{fig:hits_test} for the validation (left) and test data set (right) respectively, where in the case of the test data set the generated point cloud has been translated to the new impact position and then projected into the local cell geometry used for the full simulation with \geant. Both comparisons show equally reasonable agreement between \geant and \textsc{CaloClouds} with no visible deterioration for the translated point cloud.

\subsection{Timing}
\label{sec:Results_Timings}

\begin{table*}[h!]
\sisetup{
separate-uncertainty=true,
table-format=4.3(5)
}
\centering
\vspace{15pt}
\begin{tabular}{ll|cr}
\toprule
Hardware & Simulator &  {Time / Shower [ms]} & Speed-up \\\midrule
CPU & \geant & 4082 $\pm$ 170 & $\times 1$ \\
    & & & \\
    & \textsc{CaloClouds} & 3509 $\pm$ 220 & $\times 1.2$ \\
    & & &\\
GPU & \textsc{CaloClouds} & 38 $\pm$ 3 & $\times 107$ \\\bottomrule 
\end{tabular}
\vspace{15pt}
\caption{Comparison of the computational performance of \textsc{CaloClouds} to the baseline \geant simulator on a single core of an Intel\textsuperscript{\tiny\textregistered} Xeon\textsuperscript{\tiny\textregistered} CPU E5-2640 v4 (CPU) and an NVIDIA\textsuperscript{\tiny\textregistered} A100 with 40~GB of memory (GPU). Showers were generated with incident energy uniformly distributed between 10 and 100~GeV. The batch size was set to 1 on the CPU and to 64 on the GPU. Values presented are the means and standard deviations over 25 runs. The \geant time is taken from Ref.~\cite{gettinghigh}.}
\label{table:timing}

\end{table*}

The main motivation for simulation using generative models is speeding up the generation time per shower. In Table~\ref{table:timing}, we present the generation time of \textsc{CaloClouds} in comparison with the baseline \geant simulation on either a single CPU core or on a machine equipped with an NVIDIA A100 GPU. The \geant baseline time is taken from Ref.~\cite{gettinghigh}, which was obtained in an identical detector model and sub-detector to that studied here.
In all cases, we consider uniform energy distributions with incident particle energies between 10 and 100~GeV. 
Note that the range of 10 -- 100 GeV for timing studies is chosen for consistency with Ref.~\cite{gettinghigh} but exceeds the training range of 10 -- 90~GeV
used for the \textsc{CaloClouds} model. However, while this might in principle be out-of-distribution for modeling the data well, we expect the timing performance to be robust also in extrapolation.

On a single CPU core, the speed-up provided by \textsc{CaloClouds} is about $1.2\times$, whereas on a GPU the relative speed-up is increased to $107\times$ faster than \geant. 

When sampling, the Shower Flow is evaluated once for all events, followed by batch-wise sampling from the Latent flow and the PointWise Net. 
Unlike previous image-based fast simulation models such as Ref.~\cite{gettinghigh,  Diefenbacher:2023prl, Diefenbacher:2023angles}, the computational cost of the PointWise Net scales $\mathcal{O}(N)$ with the number of points and hence with the energy $E$ just like \geant. 
Therefore, to improve overall training and sampling speed, we batch together events with a similar number of points.
Overall, the model is trained for 800k iterations on an NVIDIA~A100, which takes about 80 hours.

Our implementation of PointWise Net uses a 100-step reverse diffusion process for shower generation. This reverse diffusion process is implemented as a computational loop and can therefore not be parallelized. 
Recent developments investigated how few-step or even single-step generation with diffusion models might be possible, e.g. using consistency distillation~\cite{song2023consistency} or progressive distillation~\cite{salimans2022progressive}, which has been already applied in high-energy physics~\cite{Mikuni:2023dvk}. 
Using one of these techniques will likely speed up the generation by at least $10\times$ since the diffusion process in our model is significantly slower than the other components such as the Shower and Latent Flow generation. Since this work is focused on model development, such  improvements are left to future work.

\section{Conclusions}
\label{sec:Conclusion}

Motivated by the overarching need of simulating particle showers in complex calorimeter detectors in a time and resource-efficient manner, generative models are widely explored in current particle physics research. 
A key problem is how to scale up such generators to be able to simulate showers with thousands of particles in high-resolution calorimeter regions. Given the inefficient scaling behavior of fixed-structure approaches, point clouds are an attractive alternative and form the foundation of our proposed strategy.

This work makes several contributions to advance the state-of-the-art in this task.
At the core of the proposed \textsc{CaloClouds} architecture is the PointWise Net, a permutation invariant, diffusion-based, point cloud generator. It is accompanied by an EPiC Encoder (in the training phase), a Latent Flow (in the sampling phase) to provide realistic conditions, and an additional Shower Flow that learns the visible energy distribution and the number of points per layer. 

In principle, this combination of four models could already be used to directly simulate a point cloud of calorimeter hits. However, the relatively dense arrangement of hits is difficult to model, with the correct number of hits per volume, in particular, being difficult to learn. Additional difficulties arise from realistic calorimeter cell layouts due to staggering between layers or gaps between sensors, resulting in different local cell topologies  in different parts of the calorimeter.
In order to circumvent these constraints we, for the first time, move to an even higher-resolution space and consider individual \geant steps. While this increases the number of points to simulate by a small integer factor, it yields a distribution that can be more easily learned and subsequently downsampled to the actual resolution of a given calorimeter.

Together these two innovations allow us, for the first time, to generate high-cardinality point clouds with unprecedented fidelity in all observed distributions with moderate speed-ups compared to \geant. 
In follow-up studies, techniques such as consistency distillation~\cite{song2023consistency} have been proven to significantly increase the sampling speed~\cite{caloclouds_2}.
Additional comparisons of the \textsc{CaloClouds} model to already established fixed-structure calorimeter generative models, i.e. the BIB-AE~\cite{gettinghigh} and CaloScore~\cite{Mikuni:2022xry}, we leave to further research.
Based on the shown results, these models currently offer higher generative fidelity, yet are geometry-dependent and suffer from the discussed computational inefficiencies.
For future research, we plan to improve the \textsc{CaloClouds} diffusion paradigm and the model architecture, taking inter-point correlations into account, to increase the generative fidelity.

For scenarios where higher resolution versions of simulations are not available for training (e.g. shower simulations, samples taken from collider data, the ongoing \textit{CaloChallenge}~\cite{CaloChallenge} competition or problems from other domains), it might be possible to generate surrogate high-resolution data by smearing points with an appropriate function. Even so, \textsc{CaloClouds} demonstrates the potential of scaling point cloud simulations with high fidelity to much higher cardinalities, making the realistic simulation of showers at the HL-LHC, the ILC, and beyond a feasible target.

\appendix

\section{Effects of the Pre-Clustering}
\label{App:Clustering}

\begin{figure*}[h]
    \centering
    \includegraphics[width=0.95\textwidth]{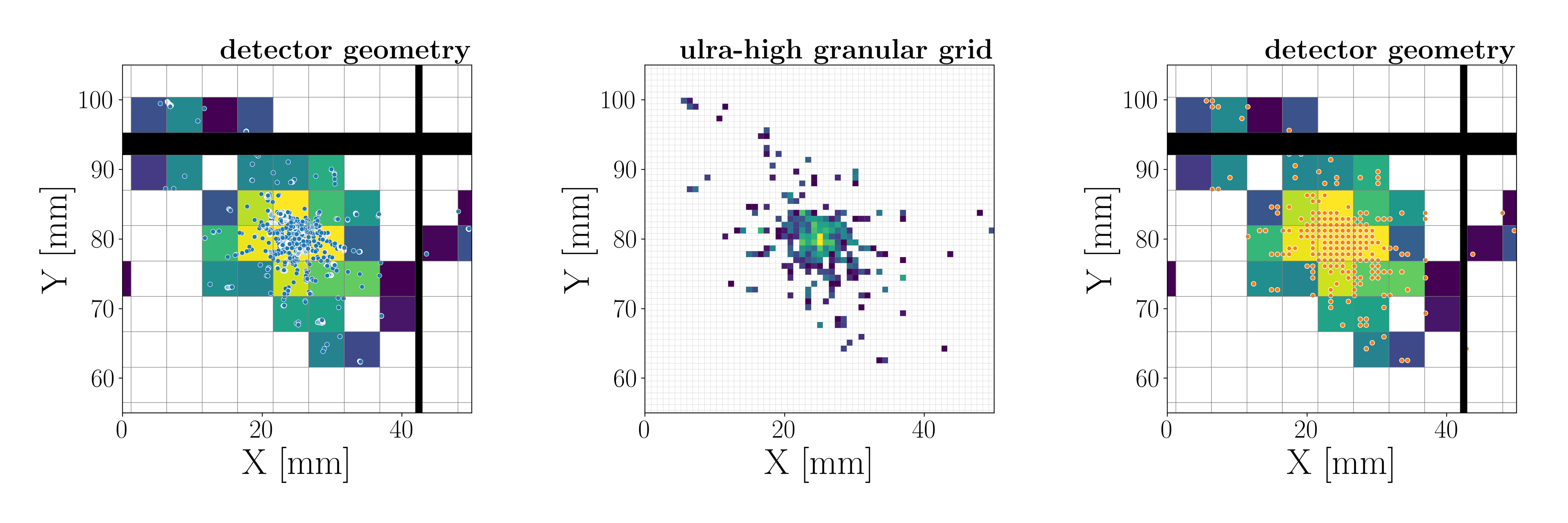}
    \caption{Example shower of 90~GeV in one layer of the calorimeter. Left: cell energies as recorded in the ILD simulation model with blue dots representing the \geant steps. Center: projection of the steps in the ultra-high granularity grid (pre-clustering). Right: Resulting cell energies after projecting the space points (orange dots) back into the detector cells.}
    \label{fig:projections_single}
    \hspace{0.5cm}
\end{figure*}

In order to study possible effects of the pre-clustering procedure (see Section~\ref{sec:Data}), we test its closure by applying the procedure to a sample simulated with \geant alone and then project the resulting clusters that form the point cloud used in \textsc{CaloCloud} back into the realistic cell geometry of the ILD simulation model. Figure~\ref{fig:projections_single} shows this for a typical shower with 90~GeV incident energy in a given layer of the calorimeter (layer 21).
The left figure shows the cell energies in this layer as generated during the \geant simulation by projecting the energy depositions into the calorimeter cells. The thick black bars are gaps between sensors as they occur in the simulation model of ILD at the chosen impact point. The center plot shows the resulting space points after the projection of the \geant steps into the virtual ultra-high granular grid. Finally, the right plot shows the cell energies resulting from simply projecting the space points back into the calorimeter cells. Additionally, the \geant steps are shown on the left in blue and the clusters on the right in orange. No obvious discrepancies are seen in this example.

\begin{figure*}[h]
    \centering
    \includegraphics[width=0.313\textwidth]{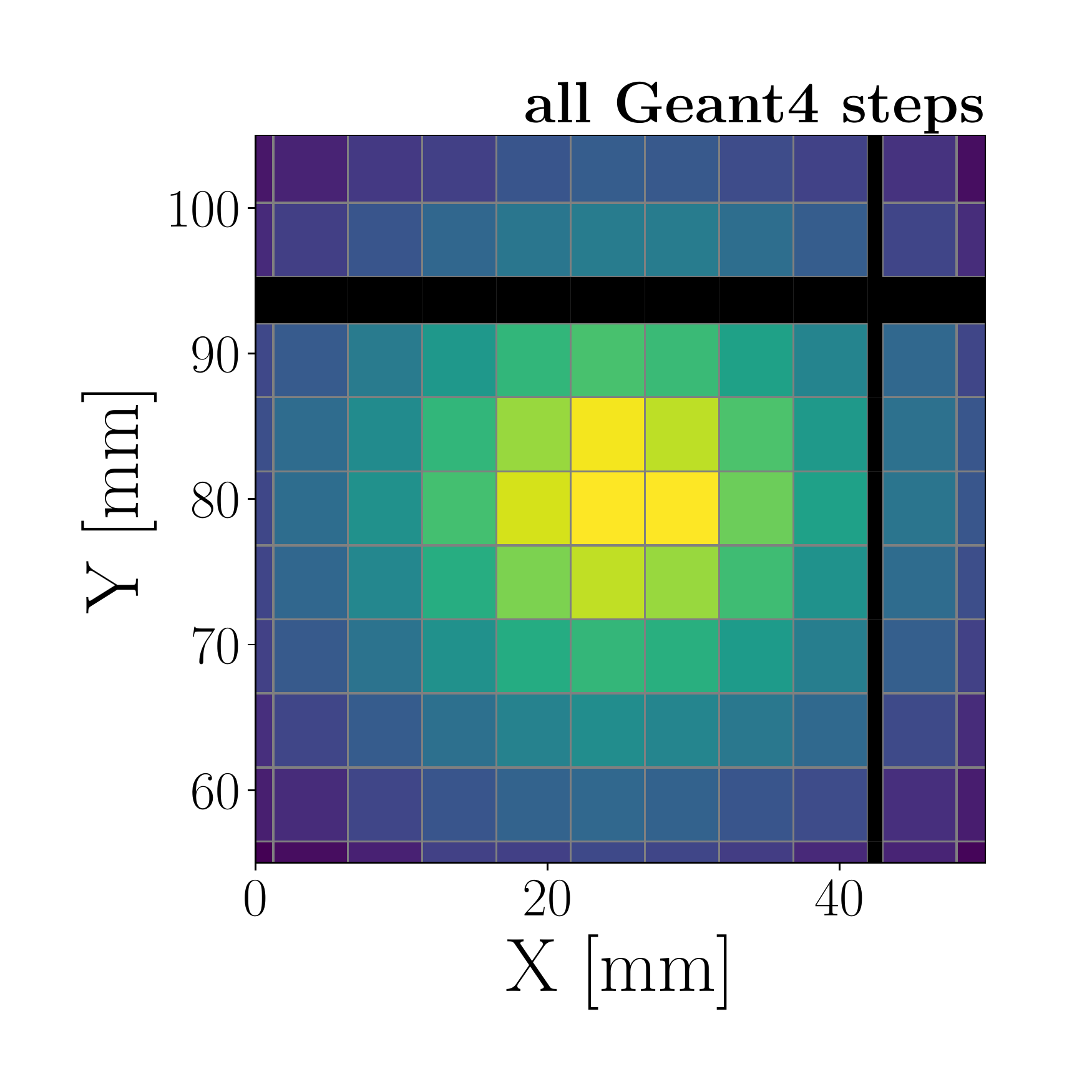}
    \includegraphics[width=0.313\textwidth]{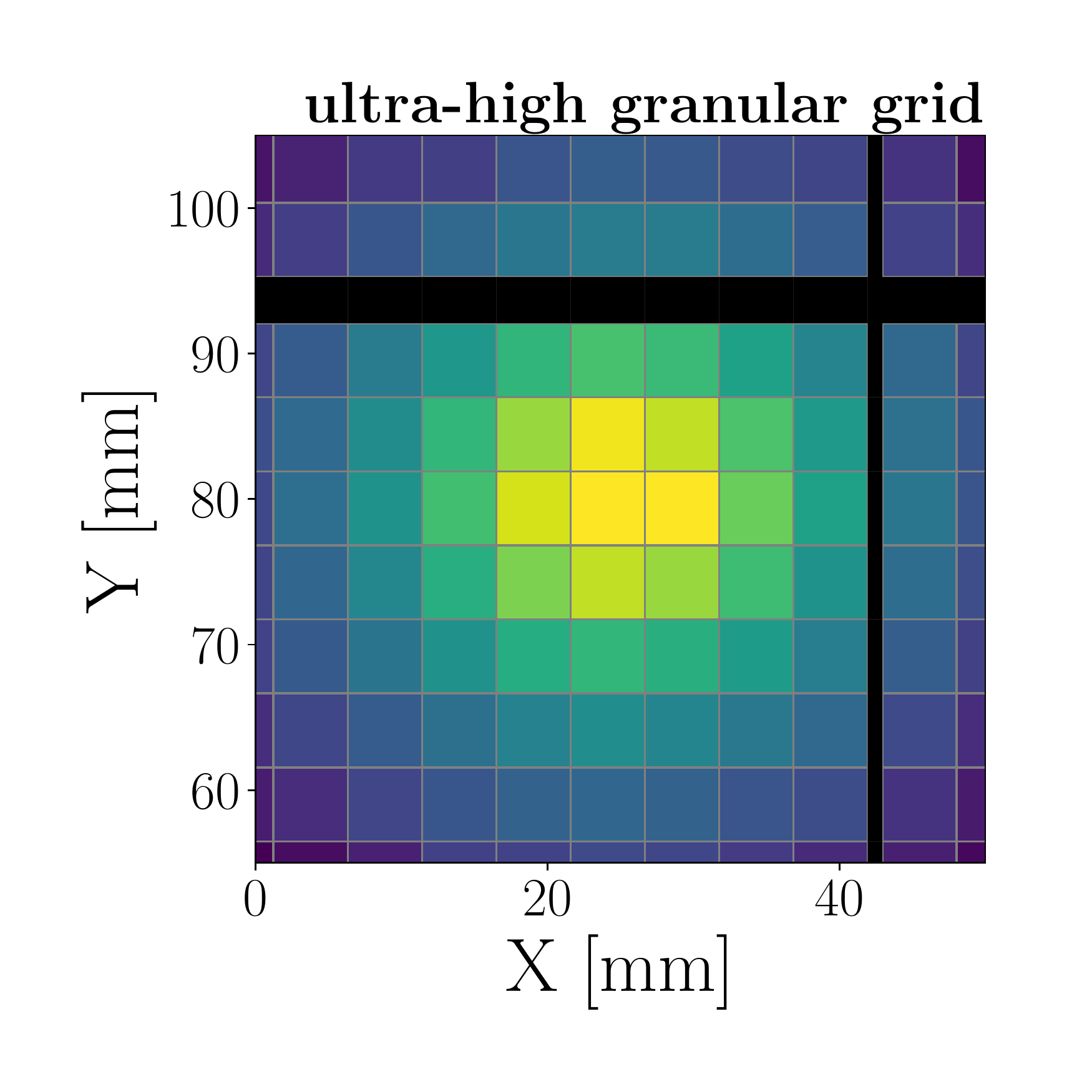}
    \includegraphics[width=0.355\textwidth]{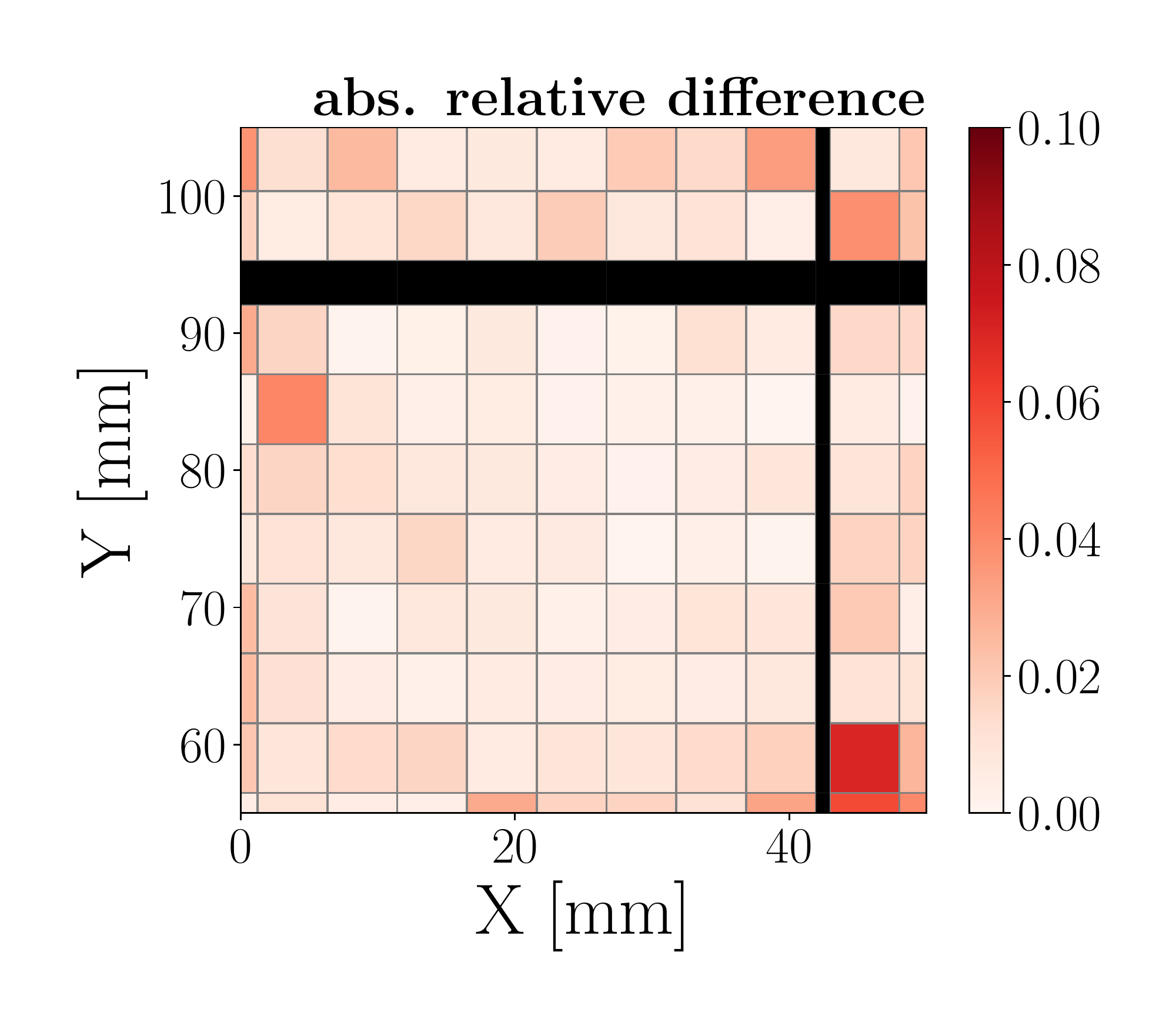}
    \caption{Overlay of 2000 90~GeV showers in one calorimeter layer. Left: as simulated in \geant. Center: same showers after a full round trip of pre-clustering and projecting back into the detector cells. Right: relative difference of the initial and processed distribution.
    }
    \label{fig:projections_overlay}
    \hspace{0.5cm}
\end{figure*}

In order to quantify the effects, in Figure~\ref{fig:projections_overlay} we show an overlay of 2,000 showers in the same layer. The left figure shows this overlay as produced by \geant, while the center plot shows this overlay after a full round trip of pre-clustering and projecting back into the same cell geometry. The right plot shows the resulting relative difference for every cell. The resulting difference is well below 10\% for all and below 2\% for most cells, indicating good closure. Importantly, most differences occur in sparsely populated parts of the showers where small differences in absolute energies lead to large relative differences. The core, where most of the shower energy is concentrated, is especially well preserved.

\begin{figure*}[h]
    \centering
    \includegraphics[width=0.355\textwidth]{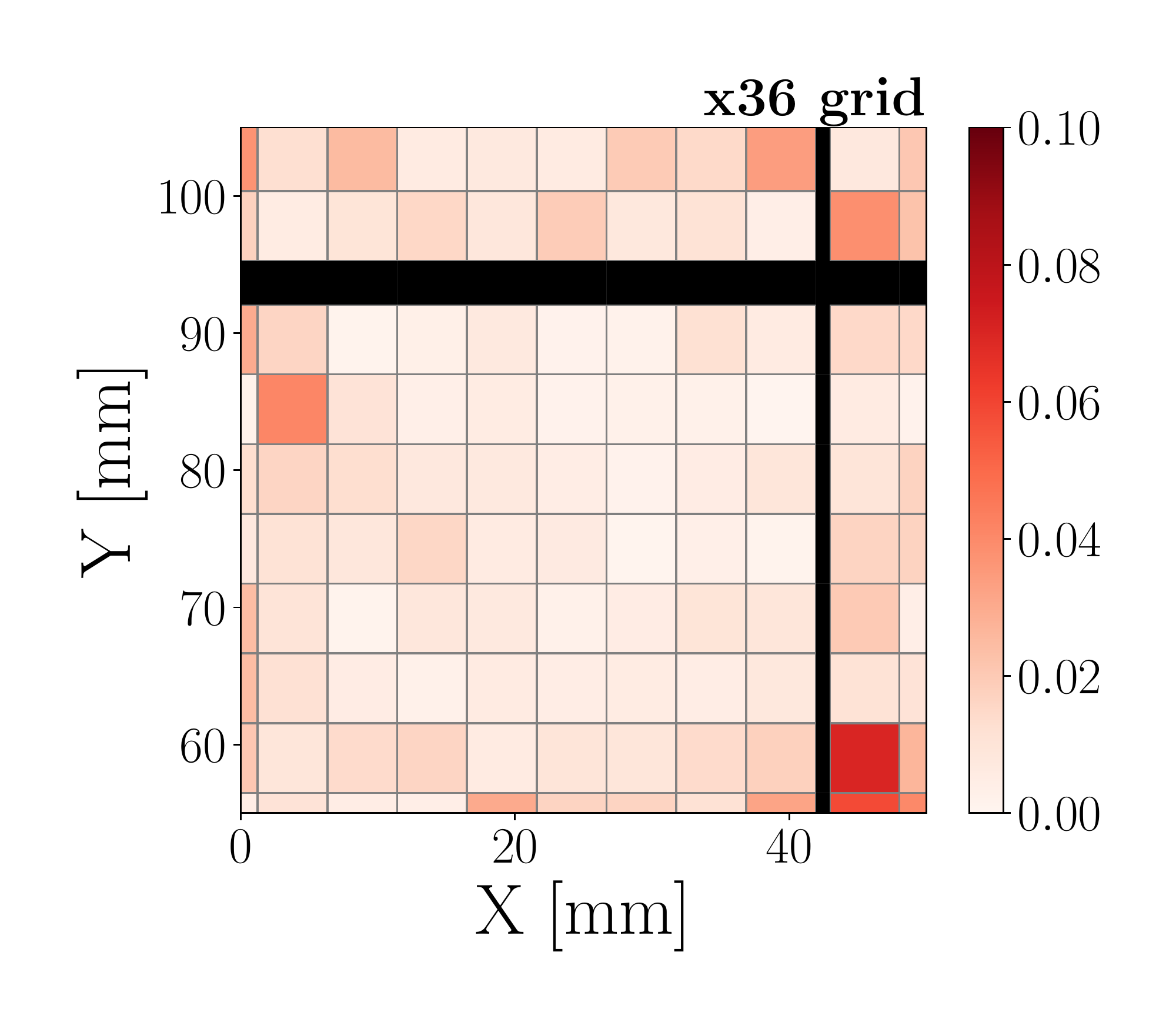}
    \includegraphics[width=0.355\textwidth]{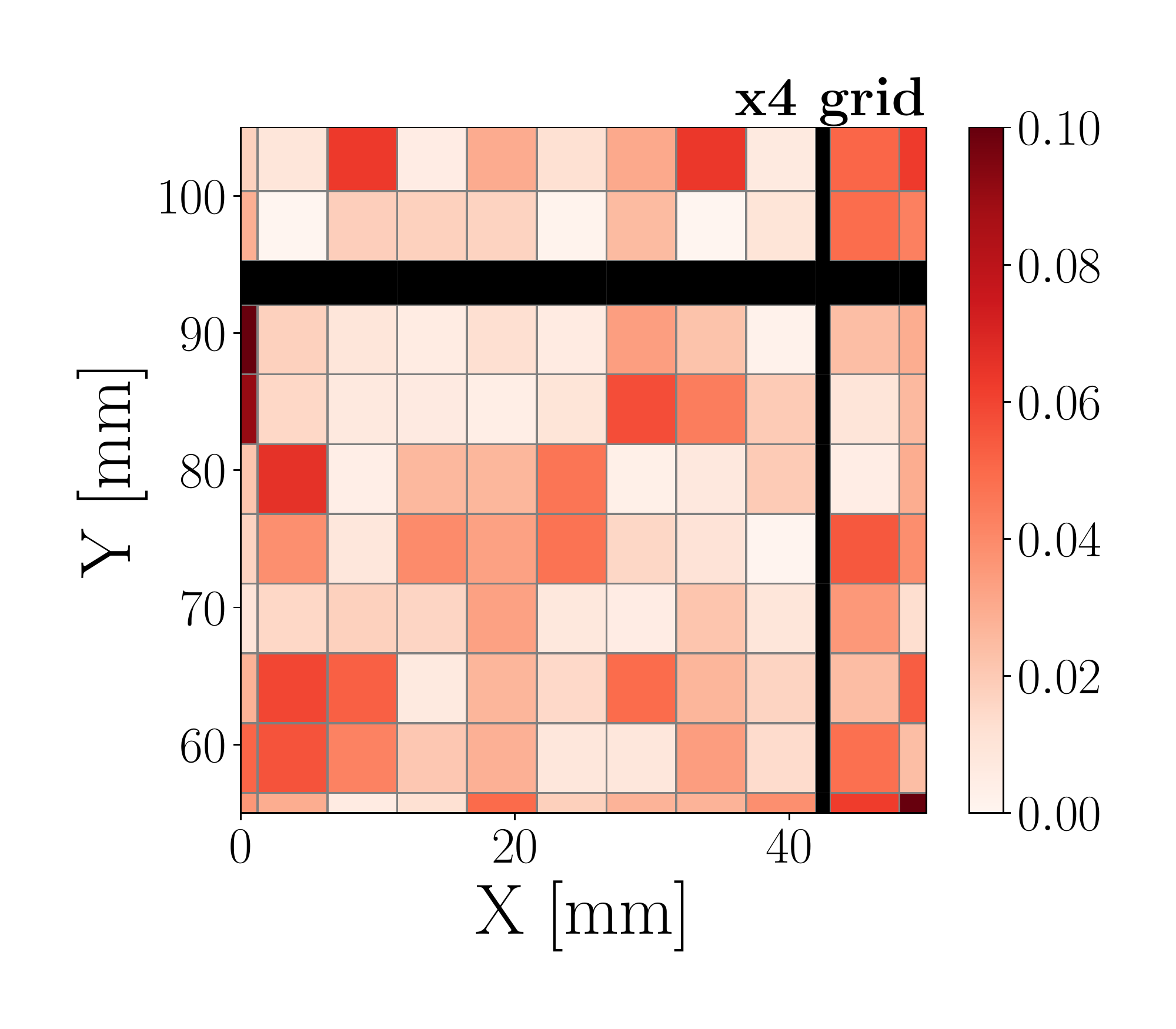}
    \caption{Difference after pre-clustering and projecting back into the cell geometry for 36 times (left) and 4 times (right) increased granularity.}
    \label{fig:projections_diff_comparison}
    \hspace{0.5cm}
\end{figure*}

Finally, Figure~\ref{fig:projections_diff_comparison} presents a comparison of this difference plot for the applied procedure with 36 times higher granularity (left - same as Figure~\ref{fig:projections_overlay}-right), as used in this paper, and a difference plot arising from a granularity increased by only 4 times (left). While such a lower granularity would potentially result in faster sampling times, it also shows significantly larger differences compared to the applied pre-clustering procedure.

\acknowledgments

We thank Sarah Heim for the valuable comments on the manuscript.
This research was supported in part by the Maxwell computational resources operated at Deutsches Elektronen-Synchrotron DESY, Hamburg, Germany. This project has received funding from the European Union’s Horizon 2020 Research and Innovation programme under Grant Agreement No 101004761. 
We acknowledge support by the Deutsche Forschungsgemeinschaft under Germany’s Excellence Strategy – EXC 2121  Quantum Universe – 390833306 
and via the KISS consortium (05D23GU4, 13D22CH5) funded by the German Federal Ministry of Education and Research BMBF in the ErUM-Data action plan.
E.B. is partially funded by a scholarship from the Friedrich Naumann Foundation for Freedom.
E.B. and W.K. acknowledge funding by the German Federal Ministry of Science and Research (BMBF) via \textit{Verbundprojekts 05H2018 - R\&D COMPUTING (Pilot\-maß\-nah\-me ErUM-Data) Innovative Digitale Technologien f\"ur die Erforschung von Universum und Materie}.
A.K. has received support from the Helmholtz Initiative and Networking Fund’s initiative for refugees as a refugee of the war in Ukraine.

\bibliographystyle{JHEP.bst}
\bibliography{main.bib}

\end{document}